\documentclass[manuscript]{emulateapj}
\usepackage{natbib}
\usepackage{color}
\usepackage{amsmath}
\shorttitle{Radio--FIR correlation}
\shortauthors{A. Basu et al.}

\begin{document}


\title{Radio--Far infrared correlation in ``blue cloud'' galaxies with $0<z<1.2$}


\author{Aritra Basu$^1$, 
        Yogesh Wadadekar$^1$, 
        Alexandre Beelen$^2$, 
        Veeresh Singh$^{2,3}$, 
        K. N. Archana$^{1,4}$, 
        Sandeep Sirothia$^{1}$, 
        C. H. Ishwara-Chandra$^1$, 
          }
\affil{$^1$National Centre for Radio Astrophysics, TIFR, Post Bag 3, Ganeshkhind, Pune - 411007, India \\
$^2$Institut d'Astrophysique Spatiale, B$\hat{\rm a}$t. 121, Universit{\'e} Paris-Sud, 91405 Orsay Cedex, France\\
$^3$Astrophysics and Cosmology Research Unit, School of Chemistry and Physics, University of KwaZulu-Natal, Durban 4041, South Africa\\
$^4$Department of Computer Science, University of Kerala, Kariavattom, Thiruvananthapuram - 695581, India\\
 }

\begin{abstract} 

We study the radio--far infrared (FIR) correlation in ``blue cloud''
galaxies chosen from the PRism MUltiobject Survey (PRIMUS) up to redshift
($z$) of 1.2 in the XMM-LSS field.  We use rest-frame emission at 1.4 GHz 
in the radio and both monochromatic (at 70$\mu$m) and bolometric (between
$8-1000~\mu$m) emission in the FIR.  To probe the nature of the correlation up
to $z\sim1.2$, where direct detection of blue star-forming galaxies is
impossible with current technology, we employ the technique of image stacking
at 0.325 and 1.4 GHz in the radio and in six infrared bands, viz. 24, 70, 160,
250, 350 and $500~\mu$m.  For comparison, we also study the correlation
for more luminous galaxies that are directly detected. The stacking analysis
allows us to probe the radio--FIR correlation for galaxies that are up to 2
orders of magnitude fainter than the ones detected directly.
The $k-$correction in the infrared wavebands is obtained by fitting the
observed spectral energy distribution (SED) with a composite mid-IR power law
and a single temperature greybody model.  We find that the radio luminosity at
1.4 GHz ($L_{\rm 1.4GHz}$) is strongly correlated with monochromatic FIR
luminosity at 70 $\mu$m ($L_{\rm 70\mu m}$) having slope $1.09\pm0.05$ and with
bolometric luminosity ($L_{\rm TIR}$) having slope $1.11\pm0.04$.  The quantity
$q_{\rm TIR} (=\log_{10}[L_{\rm TIR}/(3.75\times 10^{12} L_{\rm 1.4 GHz})])$ is
observed to decrease with redshift as $q_{\rm TIR} \propto
(1+z)^{-0.16\pm0.03}$ probably caused due to the non-linear slope of the
radio--FIR correlation. Within the uncertainties of our measurement and
the limitations of our flux-limited and color-selected sample, we do not find
any evolution of the radio--FIR correlation with redshift.

\end{abstract}


%%%%%%%%%%%%%%%%%%%%%%%%%%%%% %%%%%%%%%%%%%%%%%%%%%%%%%%%%
\section{Introduction}

The radio--far infrared (FIR) correlation is one of the tightest observed
correlations in astrophysics that connects several independent physical
parameters in the interstellar medium (ISM).  The radio luminosity and the FIR
luminosity of star-forming galaxies are observed to be correlated over five
orders of magnitude for the global scale \citep{helou85, condo92, yun01,
apple04, sarge10} with dispersion less than a factor of 2.  The radio
luminosity is typically measured at 1.4 GHz and the FIR luminosity can be both
monochromatic (at 24, 60 or 70$\mu$m) or bolometric (between 40 and 120$\mu$m
or between 8 and 1000$\mu$m).

The radio--FIR correlation is well studied for galaxies in the local universe
for several classes of galaxy morphology like spirals, ellipticals, dwarf
irregulars, etc. It is known to hold good at global \citep{condo92, yun01,
price92, wunde87} as well as at local scales (few 100 pc to few kpc) within
galaxies \citep{basu12b, dumas11, hughe06, hoern98}. At the brightest end of
FIR luminosity, the relationship is observed to hold for (ultra) luminous
infrared galaxies [(U)LIRG] and star-burst galaxies. At the faintest end it
holds in dwarf galaxies \citep{chyzy11, roych12}.

It is believed that star-formation connects the two regimes of emission.
Synchrotron (also referred to as non-thermal) emission in the radio band is
caused by acceleration of cosmic ray electrons (CREs) in the galactic magnetic
field produced by supernova explosions of massive stars. In the FIR, the
emission originates due to dust re-radiation, heated by ultraviolet (UV)
photons from massive ($\gtrsim10~\rm M_\odot$), short lived ($\sim10^6$ yrs)
stars. However, the tightness seen in the correlation needs to be explained, as
a number of independent physical quantities are responsible for the emission in
each regime like, the magnetic field, number density of CREs, energy losses of
CREs, star formation history, dust/gas density, dust absorption efficiency,
etc.  Several models have been proposed to explain the tightness seen in the
radio--FIR correlation \citep[see e.g.;][]{volk89, helou93, nikla97b}. More
recent models by \citet{lacki10} and \citet{bell03} have shown that the above
mentioned factors conspire to maintain the tightness observed for the global
radio--FIR correlation.

Observationally, it is important to assess the form of the radio--FIR
correlation at high redshifts as it might depend on the evolution of ISM
parameters with redshift ($z$) like synchrotron and inverse-Compton losses,
dust content, star formation rate, magnetic field strength and overall SED
\citep[see e.g.;][]{murph09, lacki10, schle13}. Recently, \citet{schle13}
predicted a modification of the form of the radio--FIR correlation, based on
the observed relationship between magnetic field strength and star formation
rate caused due to turbulent amplification of the magnetic field.  A breakdown
in the correlation is expected depending on the dominant energy loss mechanism
of the CREs in the radio domain, i.e., synchrotron, inverse-Compton,
bremsstrahlung and/or ionization losses.

Typical (1$\sigma$) sensitivity of most of the existing deep radio surveys are
limited only to few tens of $\mu$Jy (see e.g., \citealt{bondi03} [VLA-VVDS];
\citealt{schinn10} [VLA-COSMOS]; \citealt{hodge11} [EVLA-Stripe82], etc.).
However, a few deeper surveys exists reaching 1$\sigma$ sensitivity
$<10~\mu$Jy (see e.g., \citealt{mille13} [E-CDFS]; \citealt{morri10}
[GOODS-N]). These observations can detect normal galaxies ($L_{\rm 1.4GHz}
\sim 10^{22}~\rm W~Hz^{-1}$) up to redshift of $\sim0.2$ at 1.4 GHz with
$\gtrsim5\sigma$ sensitivity, making it difficult to study the radio--FIR
correlation for such galaxies at higher redshifts.  The correlation has been
studied for (U)LIRGs with higher luminosity ($L_{\rm 1.4GHz} \gtrsim
10^{23}~\rm W~Hz^{-1}$) up to redshifts of $\sim3$ \citep{apple04, mao11,
delmo13}. Such galaxies can have significant contamination due to AGNs and
compact nuclear starbursts.  Even in the case of relatively low optical depth,
starburst related free--free absorption can give rise to substantial
obscuration \citep{condo91b} that can affect the form of the correlation.  It
is therefore imperative to study the radio--FIR correlation for less extreme
star-forming galaxies at {\it high redshifts} where the bulk of the radio and
FIR emission originates from star formation.

In this paper, we study the properties of the radio--FIR correlation, both the
slope and the traditionally defined `$q$' parameter, for a flux limited and
color selected sample in the XMM-LSS field.  We explore the correlation for
blue star-forming galaxies up to $z\sim1.2$ employing the technique of image
stacking. Due to the inherent flux limitation of the parent sample, we detect
normal star-forming galaxies up to $z\sim0.9$ and more luminous galaxies above
that.  For comparison, we study the correlation for luminous galaxies
that are directly detected in this field up to $z\sim$ 0.95.

The paper is organized as follows: In Section 2, we describe our sample
selection and data.  We discuss the technique of image stacking at 0.325 GHz
and 1.4 GHz in the radio and at 24, 70, 160, 250, 350 and 500 $\mu$m in the FIR
and the $k-$correction method in Section 3.  We present our results in Section
4 and discuss them in Section 5. Throughout this paper, we assume a flat
$\Lambda-$CDM model with $H_0 = 70\rm~km~s^{-1}$, $\Omega_{\rm M}=0.27$ and
$\Omega_{\Lambda}=0.73$.

\begin{table*}
	\begin{centering}
  \caption{Multi waveband surveys of the XMM-LSS field.}
%\scriptsize
   \begin{tabular}{@{}lcccccc@{}}
  \hline
  Survey   & Total Area  & Resolution & 5$\sigma$ sensitivity  & $N_{\rm tot}$  & $N_{\rm PRIMUS}$ & $N_{\rm match}$\\
		  & (deg$^{2}$) & ($\arcsec \times \arcsec$) & (mJy) & & & \\
 \hline
 GMRT 0.325 GHz		  & 12    & $9.4\times7.4$ 	& 0.75 & 3929   & 894 & 111 \\
 VLA 1.4 GHz$^1$  		  & 1.3   & $5\times4$ 	   	& 0.10 & 505  & 478 & 109 \\
 SWIRE 24$\mu$m$^2$   & 10.6  & $5.6\times5.6$      & 0.45 & 24799  & 9231& 1812\\
 SWIRE 70$\mu$m$^2$   & 10.4  & $16.7\times16.7$    & 2.75  & 802   & 301 & 76\\
 SWIRE 160$\mu$m$^2$  & 10.3  & $35.2\times35.2$    & 17.5  & 286   & 106 & 28\\
 HerMES 250$\mu$m$^3$ & 18.87 & $18\times18$   	& 25.8  & 37905 & 7331& 3004\\
 HerMES 350$\mu$m$^3$ & 18.87 & $25\times25$   	& 21.2  & 42398 & 8361& 3378\\
 HerMES 500$\mu$m$^3$ & 18.87 & $37\times37$   	& 30.8  & 36933 & 7293& 2856\\

\hline 
\end{tabular}\\
\end{centering}
%\begin{tablenotes}
%\item 
$N_{\rm tot}$: Total number of catalog sources -- $^1$\citet{simps06}; $^2$Surace et al.\footnote{http://irsa.ipac.caltech.edu/data/SPITZER/SWIRE/docs/delivery\_doc\_r2\_v2.pdf}; $^3$\citet{smith12, roseb10}\\
$N_{\rm PRIMUS}$: Number of sources within the PRIMUS footprint\\
$N_{\rm match}$: Number of sources having a PRIMUS counterpart\\
%\end{tablenotes}
\label{sourcecounts}
\end{table*}

\begin{figure*}
\begin{center}
{\mbox{\includegraphics[width=16cm, angle=0]{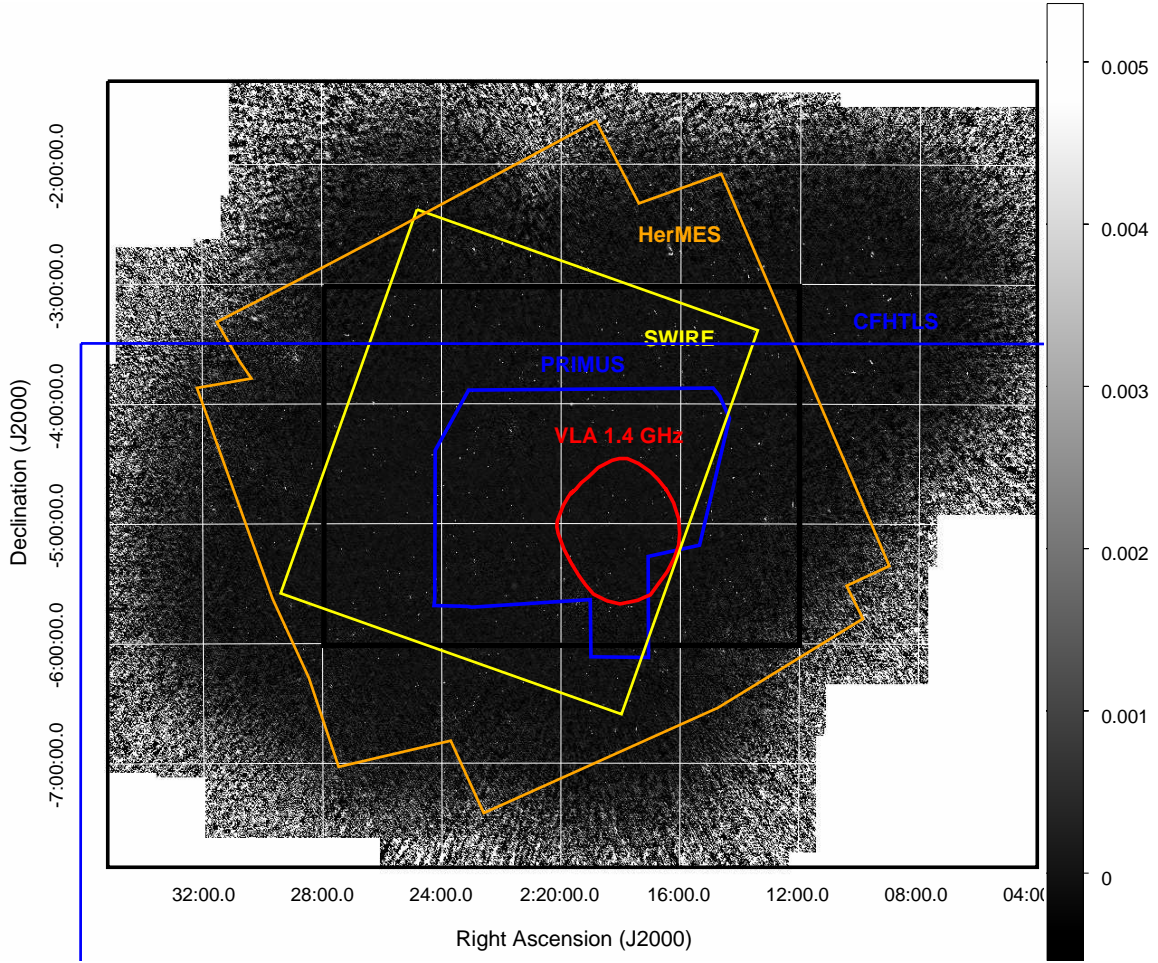}}}
\end{center}
\caption{Footprints of the various surveys of the XMM-LSS field used in
this work overlaid on the 0.325 GHz GMRT mosaic in units of Jy beam$^{-1}$. We
have saturated the image at 5 mJy beam$^{-1}$ for better representation. The
grids have a separation of 1 degree in each axis. The thick black box outlines
the 12 deg$^2$ region having smooth rms noise of $\sim150~\mu$Jy beam$^{-1}$.
The red area shows the VLA 1.4 GHz coverage of 1.3 deg$^2$ from
\citet{simps06}. The inner blue area shows the 2.88 deg$^2$ coverage of the PRIMUS
and the outer blue area shows the coverage of CFHTLS.
The yellow area outlines the SWIRE coverage of $\sim9$ deg$^2$ and the orange
area shows the HerMES coverage of the XMM-LSS field covering $\sim15$ deg$^2$.}
\label{gmrt-xmmlss}
\end{figure*}

\section{Sample selection and data}

To study the radio--FIR correlation at high redshift, one requires robust
identification of galaxies along with accurate spectroscopic redshifts. The
radio emission in galaxies mainly originates from non-thermal emission and
thermal bremsstrahlung emission. The thermal emission can affect the form of
the radio--FIR correlation \citep{hoern98, hughe06}. However, owing to the
steep spectrum, more than 95 percent of the total emission at low radio
frequencies ($\sim0.325$ GHz) is non-thermal in origin \citep{basu12a}.  Thus,
to ensure that the bulk of the radio emission is non-thermal in origin, a deep
radio survey at low frequencies, like 0.325 GHz, is necessary.  Finally, a deep
survey in the FIR regime at longer wavelengths ($\lambda \gtrsim 20~\mu$m) is
important to avoid contamination from polyaromatic hydrocarbon features in the
$5-10~\mu$m regime up to $z=1$. 

Here, we combine our Giant Metrewave Radio Telescope (GMRT) observations at
0.325 GHz with archival mid-infrared, far-infrared and 1.4 GHz radio data in
the XMM-LSS field (centered at RA=2h 21m 00s, Dec=$-4^\circ$ 30$^\prime$
00$^{\prime\prime}$ J2000).  In Table~\ref{sourcecounts} we present the salient
features of the various multiwavelength surveys used in this paper and
Figure~\ref{gmrt-xmmlss} shows the footprints of these surveys overlaid on the
0.325 GHz GMRT mosaic image.

\subsection{PRIMUS galaxy sample}

\subsubsection{Salient features of the PRIMUS survey}

Our parent galaxy sample is drawn from the PRIsm MUlti-object Survey
\citep[PRIMUS;][]{coil11}. PRIMUS is a spectroscopic faint galaxy redshift
survey to $z \sim 1$ using the Inamori Magellan Areal Camera and Spectrograph
camera on the Magellan I Baade 6.5 m telescope at the Las Campanas Observatory.
PRIMUS observed $\sim2500$ objects at once over a 0.18 deg$^2$ field of view
employing a low-dispersion prism and slitmasks.  PRIMUS has covered a total of
9.1 deg$^2$ of sky to a depth of $i_{AB} \sim 23.5$ in seven well studied
fields, one of which is the XMM-LSS field. It is in this field that the highest
number of PRIMUS spectroscopic redshifts have been obtained. The redshift
distribution of PRIMUS sample galaxies peaks at $z \sim 0.6$ and extends to $z
= 1.2$ for normal galaxies (the galaxies of interest in this paper) and $z = 5$
for broad-line active galactic nuclei.  In the XMM-LSS field, PRIMUS
observations cover $\sim2.88$ deg$^2$ sky area and detected a total of 102218
objects.  Due to the usual limitations of a prism based spectrograph, not all
redshifts are equally robust.  The PRIMUS team assigned a redshift confidence
flag $Q$ for every galaxy in their sample. They consider a redshift measurement
to be robust if $Q$ = 3 or 4. With this criterion, robust redshifts were
obtained for a total of 44451 normal galaxies and AGN in the XMM-LSS field.  We
only use the galaxies with robust redshift measurements in our analysis.  Note
that in the XMM-LSS field there are also PRIMUS calibration observations of
some sources in the VVDS field \citep{lefev05}.  This was done to verify the
quality of prism spectroscopy in PRIMUS by comparing it to higher resolution
spectroscopy done with other telescopes.  Since these calibration sources are
mostly a mix of galaxies from different spectroscopic campaigns, it is
difficult to perform any kind of statistical analysis on them.  We have
therefore excluded all sources in the calibration fields in the XMM-LSS area
from our sample.

\subsubsection{Removal of AGN}

The PRIMUS color selection is designed to select normal (non-AGN) galaxies with
$0 < z < 1.2$. Separately, a variety of targeting criteria were used to target
candidate AGN \citep[see][]{coil11}. The PRIMUS team carefully identified the
AGNs in their sample by fitting AGN spectral templates \citep[see][]{cool13}.
These are indicated by the CLASS keyword in the PRIMUS catalog. We have removed
the objects which have CLASS=AGN in our analysis.

\subsubsection{Removal of red galaxies}

A second, more serious, complication is that the PRIMUS sample contains
both ``blue cloud'' and ``red sequence'' galaxies.
These correspond to different galaxy populations with large differences in
physical parameters such as the current star-formation rate, star formation
history, stellar mass, dust content and to a large extent, galaxy morphology
\citep{tojei13}.  Broadly, the ``blue cloud'' galaxies represent active
star-forming, disk galaxies \citep{kauff03, wyder07} while the ``red sequence''
galaxies are passive and/or dust-obscured active star-forming galaxies
\citep{baldr06, taylo15}.  The significant and systematic differences in star
formation properties of these two populations are likely to impact their radio
and FIR properties as well. It is therefore important to remove the 
``red'' galaxies from our sample.

Since the radio--FIR correlation is believed to be driven by star
formation, we focus our attention on studying the ``blue cloud'' galaxies
(henceforth referred as blue galaxies).  Blue galaxies are  widely distributed
across redshifts \citep[see e.g.,][]{labbe07, bramm11}, actively forming stars 
and are supposed to dominate the cosmic star formation rate density
\citep{magne09}.  Further, they are sufficient in number to produce better
signal-to-noise ratio (S/N) for stacking analysis.  The number of red galaxies
in the PRIMUS sample are insufficient for a robust analysis and are
therefore not studied here.  We find a total of 36,776 blue galaxies in PRIMUS
at whose positions we perform image stacking in various infrared and radio
bands.

A color-magnitude diagram is used to separate ``blue'' and ``red'' galaxies.
For the PRIMUS sample, \citet{skibb14} proposed a color-magnitude diagram based
separation using the $u-g$ color and the $g-$band absolute magnitude ($M_g$).
The dividing line in the color-magnitude plot is defined as \citep{skibb14},
$$(u-g)_{\rm cut} = -0.031 M_g - 0.065z + 0.695.$$ Here, $z$ is the redshift of
the galaxy. The redshift dependence accounts for the change in the positions of
the two populations with redshift. Using this dividing line, we separate the
PRIMUS galaxies into ``blue cloud'' and ``red sequence'' galaxies.  Overall
$\sim$80 percent of the PRIMUS galaxies are blue, however, this fraction
changes slightly with redshift.

A small fraction of dusty star-forming galaxies may appear red due to reddening
and thus can be classified as red galaxies owing to the color selection. Our
sample misses star-forming galaxies where the blue light is significantly
absorbed or scattered by dust.  For PRIMUS galaxies, \citet{zhu11} showed that,
at $L\sim L_\star$, obscured star-forming galaxies comprise $\sim15$
percent of the red sequence population over all redshifts. They find, at
lower luminosities (up to 0.2$L_\star$), the fraction of obscured star-forming
galaxies in the red sequence population are up to $\sim30$ percent.  In
the PRIMUS sample, due to the inherent flux limitation, such galaxies reside
mostly in the lower redshift bins. At higher redshifts the low luminosity
galaxies are missed by the PRIMUS itself and therefore it is difficult to study
the overall fraction of obscured star-forming galaxies.  However, in our sample
of PRIMUS galaxies we find $\sim20$ percent to be red, in the redshift range
0--1.2. Thus, up to $\sim6$ percent of the entire PRIMUS sample studied here
could be dusty star-forming galaxies which are missed in our sample of blue
galaxies.

\subsubsection{Malmquist bias}

PRIMUS is a flux limited survey. It samples 100 percent of target galaxies to
$i_{AB}=22.5$ and samples galaxies sparsely up to 0.5 magnitudes fainter with
well defined a priori sampling rates. Like all flux limited surveys, it is
subject to the well known Malmquist bias. At progressively higher redshifts,
galaxies with higher intrinsic luminosities form the luminosity cutoff in the
sample.  This bias could be corrected by employing a luminosity cutoff that is
sufficiently high. Unfortunately, given our requirement for large sample size
while stacking, such an approach would not work for us. For a fixed flux limit,
and with no $k-$correction, the ratio of the minimum luminosity in the highest
redshift bin to that in the lowest one is about 125, for our chosen
cosmological parameters.  The lowest redshift bin samples galaxies that are
more than $\sim0.2$ times as luminous as the Milky Way while the highest bin
samples galaxies that are more than $\sim25$ times brighter than the Milky Way.
This implies that a true comparison of similar type of object (e.g. at a fixed
$k-$corrected luminosity or at fixed stellar mass) across redshifts is not
possible for our sample.

Further, PRIMUS is forced to exclude regions around bright stars, so there are
gaps in the spectroscopic coverage where such stars are present.  Also, the
spectrograph places limits on the minimum separation between slits.  This
implies that the highly clustered component of the sample is incompletely
sampled. 

Despite these limitations and biases, we chose to work with the PRIMUS since it
provides the largest sample of flux-limited, high-density, non-AGN galaxies
with spectroscopic redshifts over a relatively wide area to $z \sim 1.2$ that
is currently available.

\begin{figure}
\begin{center}
{\mbox{\includegraphics[width=9cm]{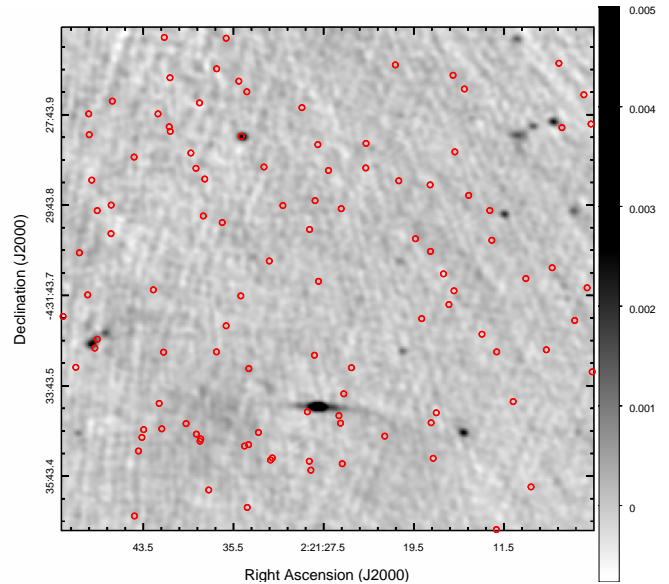}}}
\end{center}
\caption{A small portion of our 0.325 GHz image is shown with the blue PRIMUS
galaxies overplotted as red circles. It is apparent that the surface density of
PRIMUS sources far exceeds the density of radio sources.  Overall, 99.7 percent
of the PRIMUS galaxies do not have detected 0.325 GHz radio counterparts above
5$\sigma$ level.}
\label{gmrt-primus}
\end{figure}

\subsection{Data}
\subsubsection{Radio data at 0.325 GHz and 1.4 GHz}

The XMM-LSS field is observed at 0.325 GHz using the GMRT for 40 hours divided
into four 10 hour observation runs.  The field is covered by a 16-pointing
mosaic with a phase center to phase center separation of 1.0 degree. Scans are
carried out in semi-snapshot mode of 6-17 min each to optimize the {\it
uv-}coverage. This resulted in approximately uniform sensitivity within the
central region.  The final map covers the central $\sim12$ deg$^2$ having an
average 1$\sigma$ rms noise of $\sim150~\mu$Jy beam$^{-1}$ (shown as the thick
black region in Figure~\ref{gmrt-xmmlss}).  Additional area covered has higher
$rms$ noise. A detailed description of the observations and data reduction will
be presented in a forth-coming paper (Sirothia et al., in prep).

In our 0.325 GHz observations, we detected 3929 sources with signal-to-noise
ratio higher than 5$\sigma$ local rms noise within the central region of which
894 sources lie within the PRIMUS footprint. There are only 111 sources having
a PRIMUS counterpart (see Table~\ref{sourcecounts}) and 85 of them are
identified as blue galaxies.  Thus, more than 99.7 percent of the PRIMUS
galaxies are not directly detected at 0.325 GHz. The large number of 
non-detections in the radio, makes this sample ideal for stacking
analysis.  Figure~\ref{gmrt-primus} shows the positions of the PRIMUS blue
galaxies (red circles) overlaid on a small part of the 0.325 GHz radio image.
The image clearly shows that bulk of the radio sky is blank at the position of
the PRIMUS blue galaxies.

To estimate the spectral index for $k-$correcting the 0.325 GHz luminosities to
respective rest frequency at higher redshift we used the Very Large Array 1.4
GHz image from \citet{simps06} (shown as the red area in
Figure~\ref{gmrt-xmmlss}). They observed $\sim1.3$ deg$^2$ of the Subaru/{\it
XMM-Newton} Deep Field and produced a catalog of 505 sources covering the
central 0.8 deg$^2$ with peak flux density limit of $100~\mu$Jy at 5$\sigma$.
There are 478 sources lying within the PRIMUS coverage of which 109 have a
PRIMUS counterpart.

\subsubsection{FIR data between $24 - 500~\mu$m}

The Level 6 Herschel Multi-tiered Extragalactic Survey\footnote{The HerMES
project (http://hermes.sussex.ac.uk/) is a Herschel Key Programme utilizing
Guaranteed Time from the SPIRE instrument team, ESAC scientists and a mission
scientist. The HerMES data was accessed through the Herschel Database in
Marseille (HeDaM - http://hedam.lam.fr) operated by CeSAM and hosted by the
Laboratoire d'Astrophysique de Marseille.} (HerMES) in the XMM-LSS field
cover $\sim19~\rm deg^2$ having 5$\sigma$ sensitivity of about 25.8, 21.2 and
30.8 mJy at 250, 350 and $500~\mu$m, respectively using the SPIRE instrument
\citep{olive12}. The footprint of the HerMES is shown in orange in
Figure~\ref{gmrt-xmmlss}. HerMES provides images having angular resolution
$\sim18$, 25 and 37 arcsec at 250, 350 and 500 $\mu$m, respectively.  The DR2
catalog \citep{smith12, roseb10} contains 44120 sources in the XMM-LSS field
that have been detected in at least one of the SPIRE bands of which 31676
sources have been detected in all three bands.  Table~\ref{sourcecounts} lists
the number of sources detected in each of the HerMES bands along with the
number of sources within the PRIMUS coverage. Only 2445 of the PRIMUS galaxies,
i.e., about 5.5 percent, are detected in all the three bands of the HerMES.

This field is also covered by the {\it Spitzer space telescope} as a part of
Spitzer Wide-area InfraRed Extragalactic (SWIRE) survey\footnote{The Spitzer
Space Telescope is operated by the Jet Propulsion Laboratory, California
Institute of Technology, under contract with NASA. SWIRE was supported by NASA
through the Spitzer Legacy Program under contract 1407 with the Jet Propulsion
Laboratory.} \citep{lonsd03} using IRAC 3.6, 4.5, 5.8 and 8 $\mu$m bands and
MIPS 24, 70 and 160 $\mu$m bands over $\sim10~\rm deg^2$ (see
Figure~\ref{gmrt-xmmlss}). For modeling the re-processed dust emission in the
infrared bands, we combined imaging from the SWIRE\footnote{Downloaded from:
http://swire.ipac.caltech.edu/swire/astronomers/ data\_access.html} at
$\lambda= 24$, 70 and 160 $\mu$m with the HerMES DR2 at 250, 350 and 500
$\mu$m.

\section{Analysis}

\subsection{Image stacking}

Most radio images are overwhelmingly dominated by ``blank sky'' because radio
source densities on the sky are very low, at the sensitivity we can achieve
with current technology. Even in our relatively deep 0.325 GHz image we only
detect about 500 radio sources per square degree above $\sim500~\mu$Jy
($>3\sigma$). In contrast, the dense sampling in PRIMUS yields $\sim2500$
galaxies per 0.18 deg$^2$, corresponding to a source density of $\sim 14000$
deg$^{-2}$. In such a situation, stacking on the radio image at the positions
of the PRIMUS galaxies allows us to extract signals from the noise
\citep{white07}. The stacking process is straightforward -- it involves making
a cutout centered at the position of each sample galaxy and then obtaining a
mean stack image by averaging over all cutouts on a pixel-by-pixel basis. 

We note that some authors have chosen to do a median stack instead of a mean
(average) stack \citep[e.g.][]{white07}. The median has the advantage of not
being biased by a few outliers anywhere in the image.  However, the
interpretation of the median value for low S/N data typically involved in image
stacking is complicated.  For such data, simulations carried out by
\citet{white07} show that the computed median value is shifted from the true
median towards the ``local mean'' value. Here, the ``local mean'' is defined as
the mean of the values within approximately one rms of the median.
Unfortunately, the degree of the shift depends on the noise level; as the noise
increases, the recovered value approaches the local mean. Due to this effect,
the recovered median value is a function of both the intrinsic distribution of
the parameter being measured (in our case, radio flux) and the noise level.
\citet{white07} note this limitation but choose to go with the median stack
since their sample -- a large number of quasars drawn from the Sloan digital
sky survey -- has quite a few extreme outliers in the form of radio-loud
quasars. Our sample is free of extremely bright radio counterparts; in fact,
more than 99 percent of our sample galaxies have no detectable radio
counterpart at 0.325 GHz even with our deep radio imaging.  For this reason, we
chose to perform mean stacks since the statistical properties of the mean are
very well understood, even in low S/N situations.
 
We stacked the images at 0.325 and 1.4 GHz in the radio and at 24, 70, 160,
250, 350 and $500~\mu$m in the infrared at the positions of the blue galaxies.
We divided the sample into redshift bins of 0.1 between redshifts 0 to 1.2.
Table~\ref{stacks} lists the number of galaxies in each redshift bin for each
of the datasets used in this work.  Stacking was performed separately in each
redshift bin.  The PRIMUS sample has large number of galaxies in each redshift
bin with a minimum of 1100 galaxies in the bin 1.1 to 1.2 and a maximum of
$\sim5300$ galaxies in the bin 0.6 to 0.7.  The large number of galaxies in
each bin allows us to detect signal from extremely faint galaxies in the
stacked images with high signal-to-noise ratio ($\gtrsim 4$).  Due to smaller
sky coverage at 1.4 GHz, the number of PRIMUS blue galaxies available for
stacking on the 1.4 GHz  image in each redshift bin, are lower than those at
other wavebands.

Before stacking, a $61\times61$ pixel-image centered at the position of each
galaxy within the redshift range of the corresponding bin was cutout from the
images of the surveys mentioned in Section 2.  The flux density image of each
cutout was converted into a luminosity image using the respective redshift
before doing a mean stack.   We used the peak luminosity in the stacked image
at the position of the PRIMUS sources (typically the center pixel) to estimate
the mean luminosity at that redshift bin \citep[see e.g.][]{white07}.  For
unresolved point sources, the peak emission is the same as the total emission
integrated within the resolution beam.  However, this may differ if the sources
being stacked are extended w.r.t the beam \citep{karim11}. In our study, the
PRIMUS galaxies are point sources for $\sim9$ arcsec beam at 0.325 GHz.  Hence,
the peak-pixel value of the stacked imaged can be used as the integrated
luminosity of the sample. 

We carried out stacking of the flux density images also. The peak value of the
luminosity or flux density in stacked images should be computed w.r.t zero mean
background.  However, due to the contribution of emission from other sources,
the background level in the stacks is not zero mean and is slightly offset
towards positive values.  We have therefore subtracted this offset value from
all pixels in the stacked images at each redshift bin. The mean flux density,
the stacked image rms noise and the mean luminosity for each redshift bin are
tabulated in Table~\ref{stacks}. 

We note that the luminosity derived from the stacked fluxes and mean redshifts
differ from the stacked luminosities especially in the low redshift bins
(see Table~\ref{stacks}).  We list the luminosities obtained from stacking in
luminosity space (column 7) and from stacked fluxes and mean redshift (column
8). In the lower redshift bins ($z<0.3$) they differ significantly by up to
$\sim$30 percent.  However, at higher redshifts ($z>0.3$) the two methods yield
similar values within 5 percent. This is mainly because the luminosity at the
edges of the lowest $z$ bin ($0.1<z\leq0.2$) varies by a factor of 4.5 as
compared to a factor of 1.6 for a higher $z$ bin ($1<z\leq1.1$). The stacked
fluxes in a redshift bin do not account for this variation and therefore we
chose to perform stacking in luminosity space.

For robust error estimation of stacked luminosities and fluxes, we performed
standard bootstrap analysis \citep{efron94} on the cutout images in each
redshift bin.  At each redshift bin, a cutout was randomly selected $N_{\rm
obj}$ times (allowing repetitions) and were stacked. Here, $N_{\rm obj}$ is the
number of objects in a given redshift bin (see Table \ref{stacks}). We
performed 20 bootstrap iterations and computed the mean luminosity and the flux
density for each. The standard deviation for these 20 iterations is considered
as the error on the stacked quantities at a given redshift bin.

Determination of the luminosity of the stacked images could be affected if the
sources in the primary sample are clustered \citep{bethe10}.  \citet{skibb14}
found evidence of clustering of the PRIMUS galaxies, especially for the red
galaxies, which can affect the estimated stacked luminosity for images with
coarse resolution, like that of the HerMES bands.  \citet{bethe12} showed that
the contribution due to clustering can be estimated by modeling the radial
profile of the stacked image as a {\it point spread function} ({\it psf}) and
convolution of the {\it psf} and two-point angular correlation function of the
parent sample. In our case, at all the wavebands, the radial profile can be
fitted by a single component Gaussian profile within 7 percent, that represents
the {\it psf}, even at 500$\mu$m where the resolution is poorest. This
indicates that the correction due to clustering of the PRIMUS blue galaxies is
less than 10 percent and is within the errors estimated using the bootstrap
method, and hence does not significantly affect our results.  Thus,
working with blue galaxies has the advantage of being significantly less
clustered and such corrections are small enough to be ignored.

\begin{figure}
\begin{center}
\includegraphics[width=8cm]{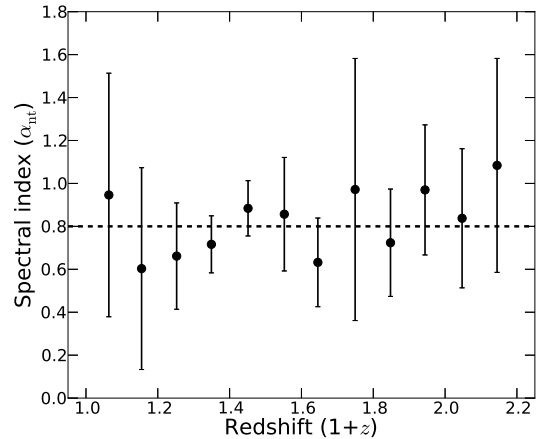}
\end{center}
\caption{Radio spectral index between 0.325 GHz and 1.4 GHz with redshift
calculated using the mean stacked luminosities. The dashed line shows the
typical spectral index of 0.8 observed for nearby normal galaxies.  The errors
are at 1$\sigma$.}
\label{alpha}
\end{figure}

\subsection{Spectral energy distribution and $k-$correction}

We obtain rest-frame emission by applying $k-$correction to the observed
luminosities in radio and FIR bands.  In this study, we explore the radio--FIR
correlation between rest-frame radio emission at 1.4 GHz and FIR emission both
monochromatic at 70$\mu$m and bolometric between 8--1000$\mu$m. These specific
radio and FIR bands are chosen to enable direct comparison with the literature
\citep[see e.g.,][]{apple04, mao11, iviso10a, iviso10b, bourn11, magne14}.  
In the radio, estimating the luminosity at a given rest frequency, in our case
1.4 GHz, is relatively easier owing to power law nature of the spectrum.
Figure~\ref{alpha} shows the variation of the spectral index, $\alpha_{\rm
nt}$, defined as $S_\nu \propto \nu^{-\alpha_{\rm nt}}$, determined using the
mean stacked luminosity at each redshift bin between 0.325 GHz and 1.4 GHz. The
spectral index in each redshift bin is consistent within $1\sigma$ errors with
typical spectral index of $\sim0.8$ (shown as dashed line in
Figure~\ref{alpha}) observed for nearby normal star-forming galaxies
\citep{nikla97a, basu12a}. We therefore estimated the rest frame 1.4 GHz
luminosity ($L_{\rm 1.4GHz, rest}$) by extrapolating the observed frame
luminosity at 0.325 GHz ($L_{\rm 0.325GHz, obs}$) assuming the typical
non-thermal spectral index of 0.8 for normal galaxies.  Thus, $L_{\rm 1.4GHz,
rest} = L_{\rm 0.325GHz, obs} \left[0.325(1+z)/1.4\right]^{0.8}$, where $z$ is
the redshift.

At FIR wavebands, due to the complex nature of the spectral energy distribution
(SED), robust modeling is necessary for accurate estimation of the
monochromatic and bolometric FIR luminosities. \citet{casey12} pointed out that
fitting a simple modified blackbody spectrum (henceforth referred to as
greybody) to the FIR spectrum is inadequate in fitting the mid-IR observation,
whereas fitting a multi temperature greybody spectrum introduces several free
parameters. Further, they demonstrated that a model consisting of single
temperature greybody + mid-IR power law fits the FIR spectrum well.
Following \citet{casey12} we modelled the FIR spectrum, $S(\lambda)$, as, 
\begin{equation}
S(\lambda) = A_{\rm GB} \frac{(1 - e^{-\tau_\lambda})\lambda^{-3}}{\left(e^{hc/\lambda kT} - 1\right)} + A_{\rm PL} \left(\frac{\lambda}{\lambda_{\rm c}} \right)^\alpha e^{-(\lambda/\lambda_{\rm c})^2}
\end{equation}
Here, $A_{\rm GB}$ and $A_{\rm PL}$ are the greybody and mid-IR power law
amplitude normalization respectively, $\lambda_{\rm c}$ is the mid-IR turnover
wavelength, $\alpha$ is the mid-IR power law index, $\tau_\lambda =
(\lambda_0/\lambda)^\beta$ is the optical depth and is unity at $\lambda_0$
(assumed to be 200 $\mu$m), $\beta$ is the dust emissivity index, $T$ is the
greybody temperature and $h, c, k$ are the Planck constant, speed of light and
Boltzmann constant, respectively. Characteristic dust temperature $(T_{\rm
dust})$ is given by Wien's displacement law, $T_{\rm dust} = b/\lambda_{\rm
peak}(\rm \mu m)$, where, $b=2.898\times10^3\rm \mu m~K$ and $\lambda_{\rm
peak}$ is the peak wavelength of the fitted SED.

\begin{figure}
\begin{center}
\begin{tabular}{c}
{\mbox{\includegraphics[height=7.5cm]{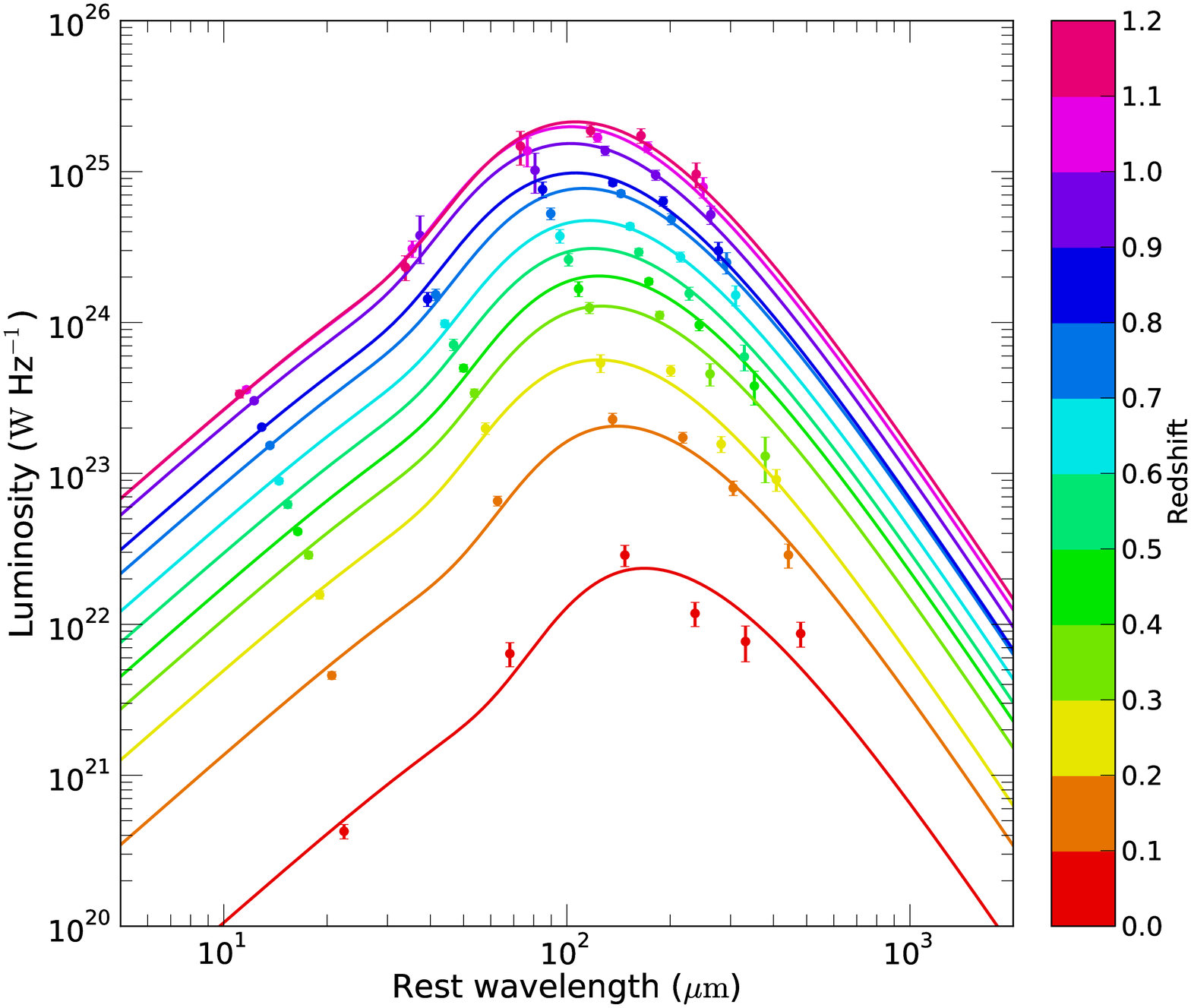}}}\\
\end{tabular}
\end{center}
\caption{The fitted SED of the stacked sources in infrared bands at various
redshift bins shown in colors. The SEDs are fitted jointly with a mid-IR power
law and a single temperature greybody.}
\label{stack-sed}
\end{figure}

Although there are six fitting parameters, namely, $A_{\rm GB}$, $A_{\rm PL}$,
$\alpha$, $\beta$, $T$ and $\lambda_{\rm c}$, the parameters $A_{\rm PL}$ and
$\lambda_{\rm c}$ are coupled to the rest, thereby reducing the number of free
parameters to four \citep[see][]{casey12}. Thus, for a robust fitting of the
FIR SED, at least 5 data points are necessary. We have therefore stacked, in
luminosity space, 6 wavelength bands in the FIR, i.e., three MIPS bands (24, 70
and $160~\mu$m) from SWIRE and three SPIRE bands (250, 350 and $500~\mu$m) from
HerMES at the position of the PRIMUS galaxies to constrain the SED. However, to
constrain the mid-IR power law index, $\alpha$, at least 3 photometric points
in the mid-IR (between $\sim10-50~\mu$m, rest-wavelength) are required, which
are not available in our case. We have therefore fixed $\alpha=2$
\citep{casey12}. Similarly, we fixed $\beta=1.5$ \citep[following][]{magne14,
casey12}.

Separate fits were done for the stacked SEDs in luminosity space, in each of
the 12 redshift bins shown in Figure~\ref{stack-sed}. Each fit is color coded
for the redshift bin. In Figure~\ref{stack-sed-comp} we show the two
components, greybody (red dashed curve) and mid-IR power law (blue dash-dot
curve) at each redshift bin along with reduced chi square ($\chi^2_{\rm
red}$) and errors in the fitted parameters.  The two components together give
an excellent fit to the data with one exception. The SED fit for the first
redshift bin is poor having large $\chi^2_{\rm red}$ and more than 50 percent
error in the derived FIR luminosity. We have excluded this bin from all our
further analysis.

\begin{figure*}
\begin{center}
\begin{tabular}{cc}
{\mbox{\includegraphics[width=9cm]{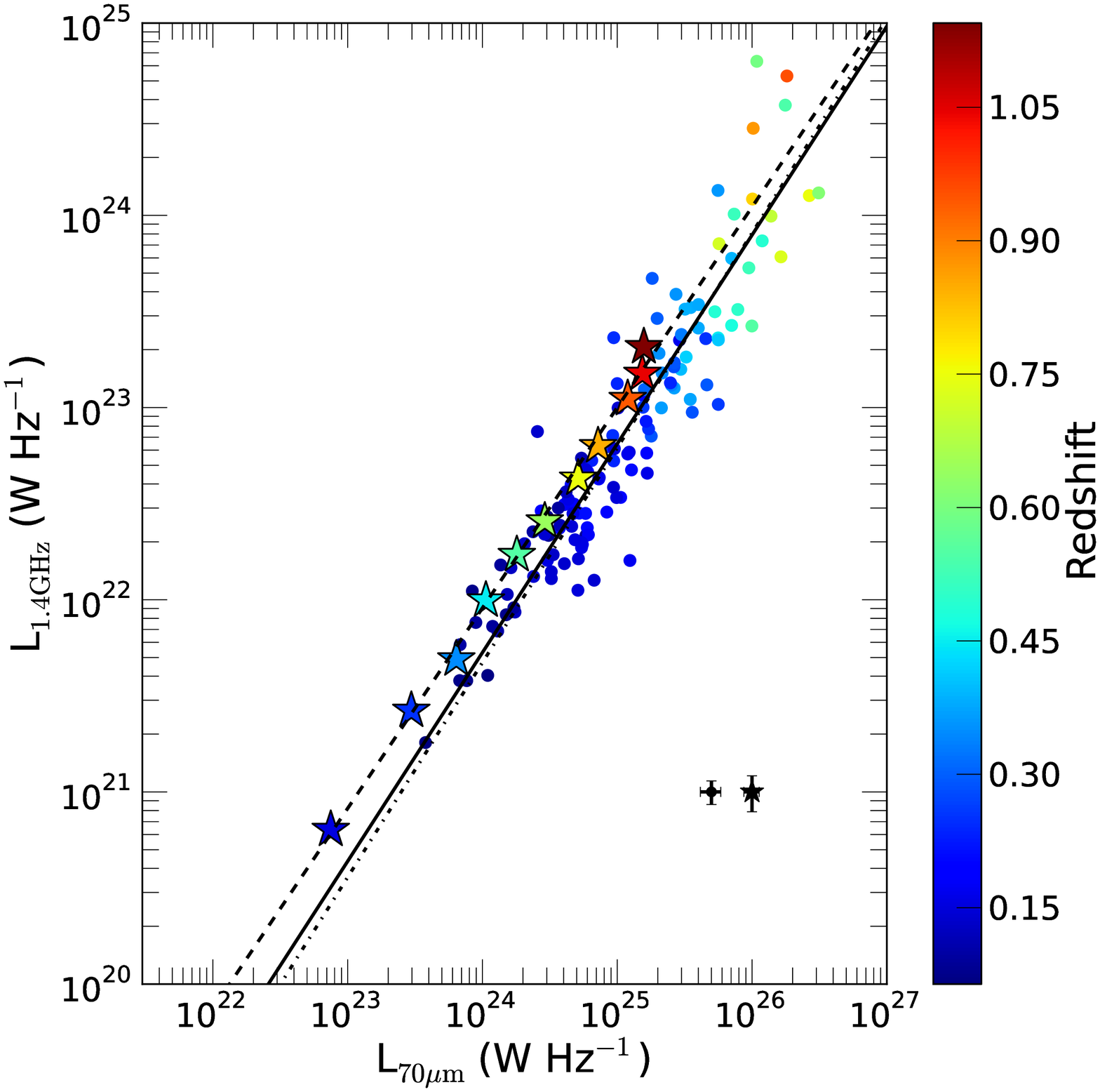}}}&
{\mbox{\includegraphics[width=9cm]{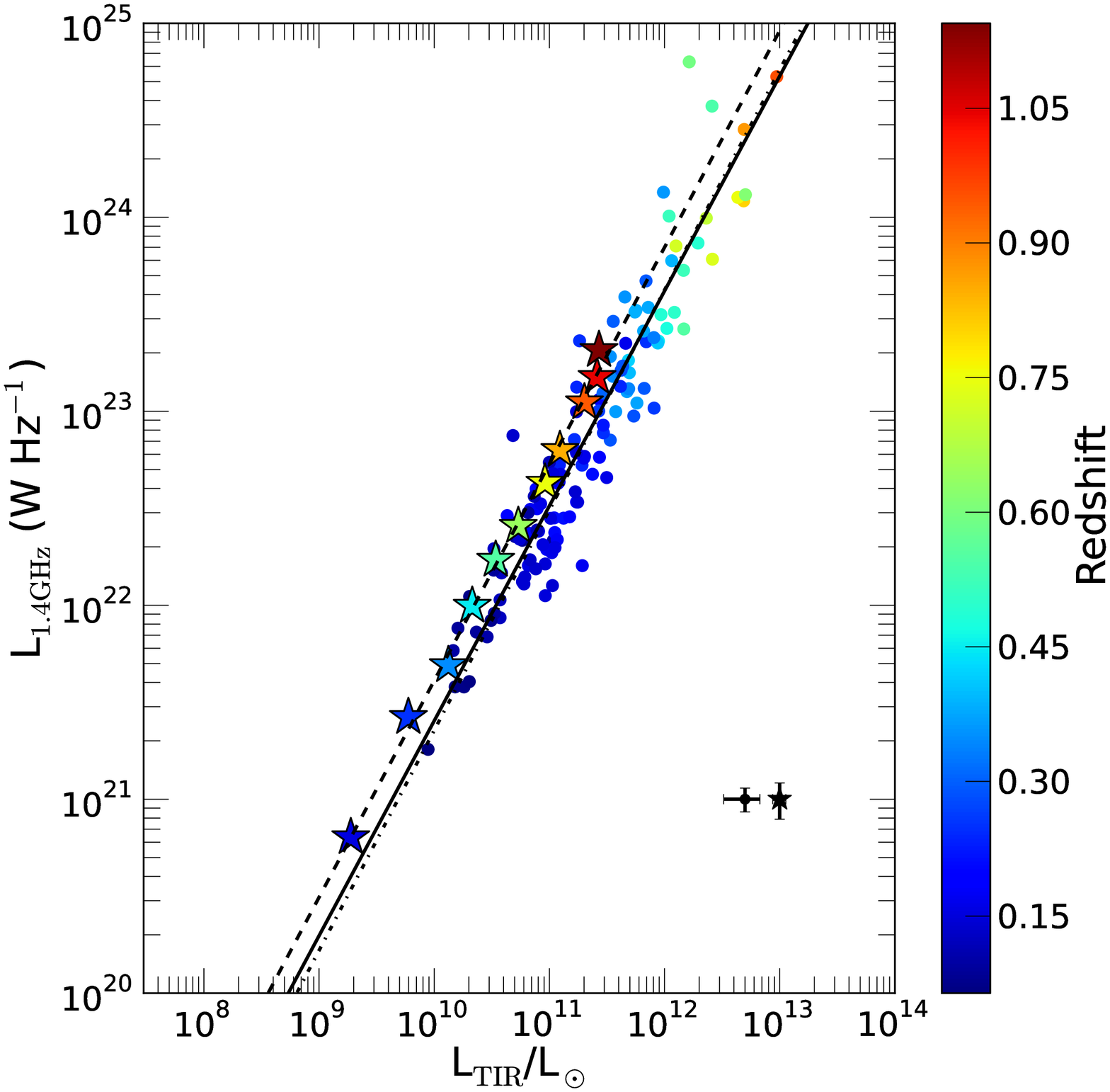}}}\\
\end{tabular}
\end{center}
\caption{{\it Left hand panel:} Luminosity at 1.4 GHz ($L_{\rm
1.4GHz}$) versus monochromatic luminosity at 70$\mu$m ($L_{\rm 70\mu m}$) at
rest-frames. The stacked sources are shown as stars and the sources detected in
the XMM-LSS field are shown as circles. The symbols are color coded based on
their redshift. The solid line shows the fit to the entire data, the dashed and
dashed-dot lines are for the stacked and detected sources respectively. The
slopes are listed in Table~2. Typical errors on the data are shown in the lower
right.  {\it Right hand panel:} $L_{\rm 1.4 GHz}$ versus bolometric luminosity
($L_{\rm TIR}$).  The symbols and the lines have the same meaning as in the
left hand panel.}
\label{frc}
\end{figure*}

To compare our results using stacking, and study its nature in perspective
of the large span in luminosity ($\sim5$ orders of magnitude), we also study the
radio--FIR correlation using detected sources along with the stacked sources.
For the sources that are detected at 0.325 GHz, we cross matched them with the
HerMES DR2 catalog \citep{roseb10}. There are 1805 sources that are
detected at 0.325 GHz and in all the three HerMES bands (250, 350 and
500$\mu$m). The HerMES bands are required for a better constraint on the ``cold
dust'' greybody spectrum. However, we note that for the above SED fitting
method to work best in estimating the 70$\mu$m monochromatic luminosity, one
needs mid-IR observation to constrain the power law component. We therefore,
additionally imposed the condition, that the sources be detected at 24 and
$70~\mu$m from the SWIRE survey. There are 231 sources detected in five
infrared bands from 24$\mu$m to $500\mu$m with 0.325 GHz counterparts. Only 26
of the 231 sources have spectroscopic redshift from the PRIMUS,
predominantly due to small areal coverage. Therefore, to
increase the sample size we used photometric redshifts ($z_{\rm phot}$) from
the Canada-France-Hawaii Telescope Legacy Survey\footnote{The CFHTLS public
data release includes observations obtained with MegaPrime/MegaCam, a joint
project of CFHT and CEA/DAPNIA, at the Canada-France-Hawaii Telescope (CFHT)
which is operated by the National Research Council (NRC) of Canada, the
Institut National des Science de l'Univers of the Centre National de la
Recherche Scientifique (CNRS) of France, and the University of Hawaii.  The
data products are produced at TERAPIX and the Canadian Astronomy Data Centre, a
collaborative project of NRC and CNRS.} \citep[CFHTLS;][]{ilber06} available in
the XMM-LSS field that covers a larger area (see
Figure~\ref{gmrt-xmmlss}). CFHTLS provides accurate $z_{\rm phot}$ for
galaxies, with only $\sim4$ percent catastrophic failures, i.e., $\Delta
z/(1+z_{\rm spec}) > 0.15$ \citep{ilber06}.  Of the 231 sources, 126 have
$z_{\rm phot}$ available.  These 126 sources are an assortment of galaxies
comprising of 100 blue and 26 red galaxies spanning up to a redshift of 0.95.
Note that, 19 of these 100 blue galaxies lie in the PRIMUS coverage and have
also been included while stacking. 

The choice of $\alpha$ plays a crucial role in determining the $A_{\rm PL}$ and
$\lambda_{\rm c}$ and, in turn the SED fit below $\sim100~\mu$m, leading to
deviant fits. We have therefore performed a grid search for the value of
$\alpha$ between 1.2 and 3.2 in steps of 0.05. We fixed $\alpha$ where the
$\chi^2$ to the fit was minimum.  More than 90 percent of the objects could be
fitted with $\alpha$ in the range 1.8--2.3.

We would like to emphasize that, the three HerMES data points help in
constraining the ``cold dust'' greybody spectrum while for the mid-IR power law
part, more than one data point below $\sim100~\mu$m observed-frame wavelength
(in our case 24 and $70~\mu$m) is necessary for robust estimation of
$k-$corrected 70$\mu$m luminosity. Using only 24$\mu$m and three HerMES bands,
constrains the FIR greybody well, but the mid-IR SED is not well constrained
leading to increased dispersion in the radio--FIR correlation when studied at
70$\mu$m.  Having the observed frame 70$\mu$m data point is crucial to our
analysis.

\begin{figure*}
\begin{center}
\begin{tabular}{cc}
{\mbox{\includegraphics[width=9cm]{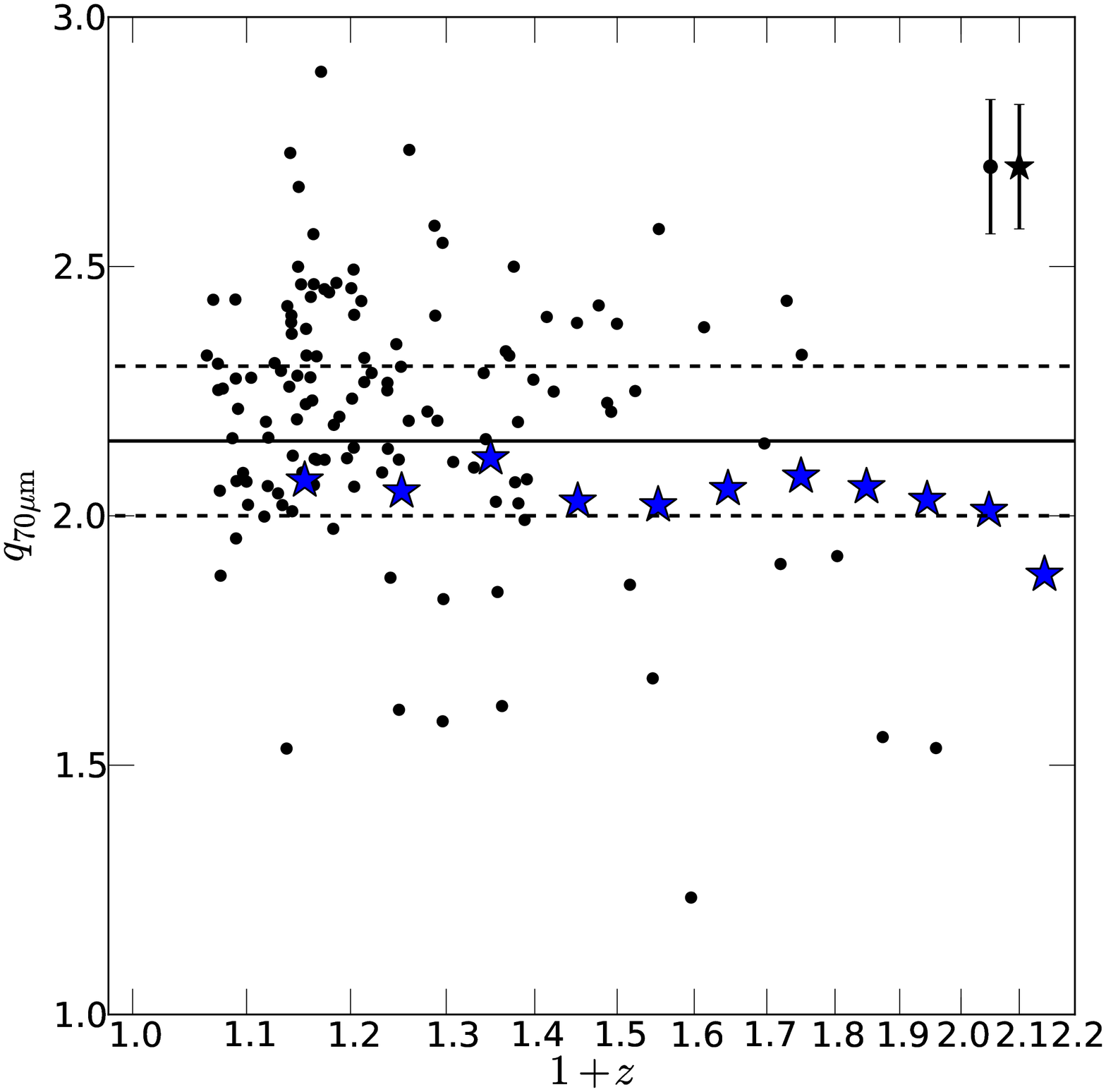}}}&
{\mbox{\includegraphics[width=9cm]{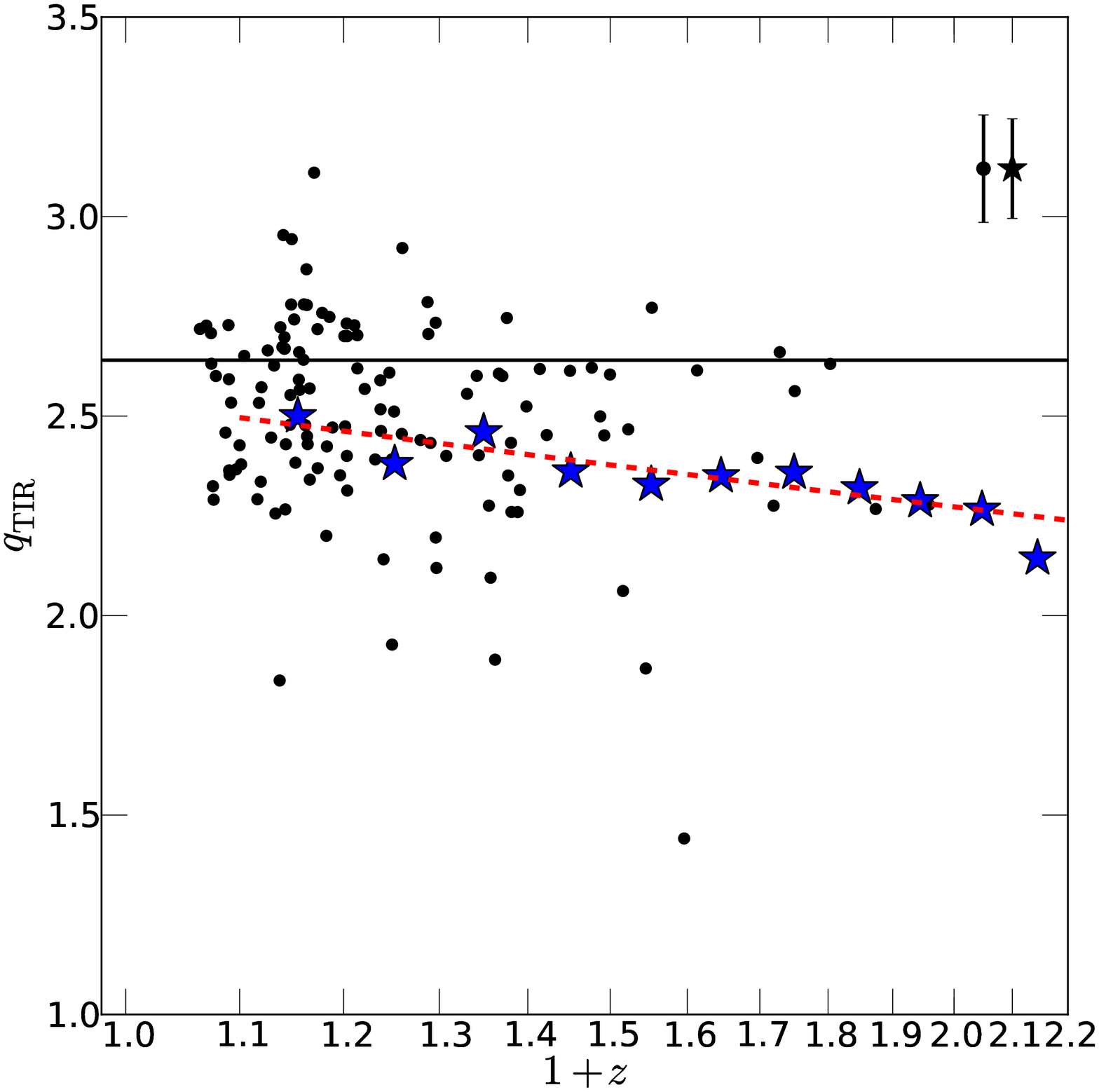}}}\\
\end{tabular}
\end{center}
\caption{{\it Left hand panel:} The variation of $q_{\rm 70\mu m}$ with
redshift ($1+z$). The stars and circles represent stacked and detected sources
respectively. The solid and the dashed lines shows the mean value of $q_{\rm
70\mu m}$ and 1$\sigma$ dispersion respectively from \citet{apple04}.  {\it
Right hand panel:} The variation of $q_{\rm FIR}$ with redshift ($1+z$). 
The red dashed line shows the fitted curve for the stacked
sources of the form $(2.53\pm0.04) (1+z)^{-0.16\pm0.03}$. The solid line shows
the mean value of $q_{\rm TIR}$ from \citet{bell03} observed for local
galaxies. The mean values of these quantities obtained in our study are
given in Table~2.}
\label{q}
\end{figure*}

\section{Results}
\label{results}

Image stacking technique allows us to estimate mean luminosities of our sample
galaxies in each redshift bin of 0.1 across $0 < z < 1.2$ (see Section~3). We
summarize our results of the stacked images at various wavebands in
Table~\ref{stacks}.  Note that, we detect emission from faint galaxies that
have $L_{\rm 1.4GHz}$ of few times $\sim 10^{20}$ W~Hz$^{-1}$ and $L_{\rm 70\mu
m}$ of few times $10^{22}$ W~Hz$^{-1}$.  Using stacking we probe radio--FIR
correlation for galaxies that are up to 2 order to magnitude fainter than the
samples used in previous studies based on direct detections in the similar
redshift range \citep[see e.g.][]{garre02,apple04, kovac06, mao11, delmo13}.
However, this increased depth is restricted to the lowest redshift bins, due to
the Malmquist bias present in the parent optical sample.  The radio/FIR
stacking process does not introduce any additional biases, but it cannot
overcome the inherent bias in the parent optical sample.

The radio--FIR correlation is generally quantified by two parameters -- 1) the
slope in log-log space, $b$, given by, $L_{\rm radio} \propto L_{\rm IR}^b$ and
2) the parameter `$q$' defined as $q = \log_{10}(L_{\rm IR}/L_{\rm radio})$.
Here, $L_{\rm radio}$ is the luminosity at a radio frequency, in our case at
1.4 GHz rest-frequency. However, several definitions of $L_{\rm IR}$ are found
in the literature, such as rest-frame monochromatic luminosity at some
specified infrared wavelength \citep[see e.g.][]{apple04, mao11} or bolometric
luminosity integrated between $40-120~\mu$m \citep[defined by][]{helou85} or
between $8-1000~\mu$m \citep[see e.g][]{bell03, iviso10b, bourn11, magne14}.
Here we study the radio--FIR correlation using both rest-frame monochromatic
infrared luminosity at 70$\mu$m ($L_{\rm 70\mu m}$) as well as bolometric total
infrared luminosity integrated between $8-1000~\mu$m ($L_{\rm TIR}$).  For the
monochromatic case, we define $q_{\rm 70\mu m}=\log_{10}(L_{\rm 70\mu m}/L_{\rm
1.4GHz})$ and for the bolometric case, we define,
\begin{equation}
	q_{\rm TIR}=\log_{10}\left[\frac{L_{\rm TIR} ({\rm W})}{3.75\times 10^{12}}\right] - \log_{10}[L_{\rm 1.4GHz} ({\rm
W~Hz^{-1}})]
\label{def-q}
\end{equation}
following \citet{helou85}.  Note that \citet{helou85} defined the
bolometric infrared luminosity integrated between $40-120~\mu$m.
\citet{bourn11} showed that the use of $8-1000~\mu$m luminosities increases the
absolute value of $q_{\rm TIR}$ by $\sim0.32$ compared to that originally
defined by \citet{helou85}.  The rest-frame far-infrared monochromatic
luminosity at 70$\mu$m and the bolometric luminosity are computed using the
SED fits discussed in Section~3.2.

In Figure~\ref{frc} (left hand panel) we plot the radio luminosity at rest
frame 1.4 GHz ($L_{\rm 1.4GHz}$) vs. the FIR luminosity at 70$\mu$m ($L_{\rm
70\mu m}$) for the stacked galaxies (shown as stars) and the directly detected
galaxies (shown as circles).  The symbols are color coded based on the
redshift. We note that the mean stacked luminosities in the higher redshift
bins are higher. This trend is expected due to the inherent Malmquist bias.
The radio and FIR luminosities are found to be strongly correlated with
Spearman's rank correlation, $r>0.99$.  We fitted for the slope of the
correlation using ordinary least-square ``bisector method'' \citep{isobe90} in
log--log plane.  The solid line shows the fit to all the data points and have a
slope 1.09$\pm$0.05.  We also did separate fits for the stacked and detected
galaxies. The dashed line shows the fit for the stacked galaxies for which the
slope is found to be 1.04$\pm$0.03.  The dash-dot line represents the fit for
detected galaxies and have a slope of 1.12$\pm$0.05.  They all agree within
errors suggesting the radio--FIR correlation holds over the entire luminosity
range of five orders of magnitude, probing galaxies with different ISM
properties. The values of the slopes are tabulated in Table~2.

Figure~\ref{frc} (right hand panel) shows the radio--FIR correlation between
$L_{\rm 1.4GHz}$ and the total infrared luminosity ($L_{\rm TIR}$).  We find
strong correlation between the two quantities.  In this case, we find the slope
to be significantly steeper than unity. The slope of the radio--FIR correlation
is found to be $1.11\pm0.04$ when fitted for the stacked sources and detected
sources together (solid line). By separately fitting the stacked galaxies and
the detected sources we find the slopes to be $1.12\pm0.03$ and $1.13\pm0.04$,
respectively (see Table~2). Similar non-linear slope of $1.10\pm0.04$ was also
reported by \citet{bell03} for normal star-forming galaxies in the local
universe having $L_{\rm TIR}$ in the range $10^{8}$ to $10^{12}~L_\odot$.

In Figure~\ref{q}, we study the variation of $q_{\rm 70\mu m}$ (left hand
panel) and $q_{\rm TIR}$ (right hand panel) with redshift $(1+z)$. The mean
values of the parameter are tabulated in Table~2. We find the mean value of
$q_{\rm 70\mu m}$ to be 2.18 with 1$\sigma$ dispersion of 0.26. This is
consistent with $\langle q_{\rm 70\mu m}\rangle=2.15$ (shown as the solid line)
and 1$\sigma$ dispersion of 0.15 (shown as the dashed lines) observed by
\citet{apple04} over similar redshift range and by \citet{mao11} for redshifts
up to $\sim3$ for more luminous galaxies.  However, for the stacked sources,
$q_{\rm 70\mu m}$ is found to be lower than that observed by \citet{apple04}
and have $\langle q_{\rm 70\mu m}\rangle$ of $2.0\pm0.2$.  Within errors,
$q_{\rm 70\mu m}$ is observed to remain constant with redshift.

\begin{table*}
	\begin{centering}
  \caption{Results of the radio--FIR correlation.}
%\scriptsize
   \begin{tabular}{@{}lcccccc@{}}
  \hline
  \multicolumn{1}{l}{}  &
\multicolumn{2}{c}{$70\mu m$}  &
\multicolumn{2}{c}{TIR}  &
\multicolumn{2}{c}{FIR}  \\
		   & Slope & $\langle q_{\rm 70 \mu m}\rangle$ & Slope & $\langle q_{\rm TIR}\rangle$ & Slope & $\langle q_{\rm FIR}\rangle$\\
 \hline
 Stacks	  	    & 1.04$\pm$0.03  &  2.0$\pm$0.3	& 1.12$\pm$0.03 & 2.34$\pm$0.22   & 1.07$\pm$0.03 & 2.07$\pm$0.21 \\
 Detections (blue) & 1.05$\pm$0.05 &  2.22$\pm$0.27 & 1.12$\pm$0.06 &  2.51$\pm$0.24  &  1.07$\pm$0.06 & 2.25$\pm$0.24\\
 Detections (red)  & 1.30$\pm$0.13 &   2.16$\pm$0.40 &  1.14$\pm$0.07 &  2.48$\pm$0.32  &  1.29$\pm$0.12 &  2.19$\pm$0.31\\
 Detections (blue+red)& $1.12\pm0.05$  & $2.22\pm0.24$    & $1.13\pm0.04$	& $2.51\pm0.24$  & $1.13\pm0.05$   & $2.25\pm0.24$  \\
 Stacks+detections  & $1.09\pm0.05$  & $2.18\pm0.26$    & $1.11\pm0.04$ & $2.50\pm0.24$ & $1.10\pm0.04$ & $2.23\pm0.25$\\

\hline 
\end{tabular}\\
%\begin{tablenotes}

%\item 
The slopes are estimated by fitting a straight line in the log-log space
of the luminosity plots shown in Figure~\ref{frc}.	\\ $\langle...\rangle$
represents the mean values of the quantity along with their measurement errors
from Figure~\ref{q}. 
%\end{tablenotes}
\end{centering}
\label{tab-frc}
\end{table*}

$q_{\rm TIR}$ shows a mean value of 2.50 with a $1\sigma$ dispersion of 0.24.
This is close to the value 2.64 observed for local galaxies \citep{bell03} with
similar dispersion. The mean value of $q_{\rm TIR}$ for the stacked sources are
found to be lower, $\langle q_{\rm TIR}\rangle = 2.34\pm0.22$, than that of the
detected sources, $\langle q_{\rm TIR}\rangle = 2.51\pm0.24$ (see Table~2). Our
estimated value of $\langle q_{\rm TIR} \rangle$ for stacked sources is close
to the value $2.40\pm0.29$ observed by \citet{iviso10a} for sub-mJy radio
emitting galaxies.  Note that, the radio--FIR correlation is tighter when
studied for bolometric FIR luminosity than for monochromatic 70$\mu$m
luminosity.  This is evident from the mean value of $q_{\rm TIR}$ showing
lesser spread around the mean (0.24 dex) as compared to $q_{\rm 70\mu m}$ (0.26
dex). In linear space this corresponds to $\sim16$ percent spread in $L_{\rm
1.4 GHz}$ vs.  $L_{\rm TIR}$ as compared to $\sim28$ percent spread when
studied between $L_{\rm 1.4 GHz}$ vs. $L_{\rm 70\mu m}$.

For completeness with the various definitions of the parameter `$q$', we also
present the results of the radio--FIR correlation between rest-frame $L_{\rm
1.4 GHz}$ and $L_{\rm FIR}$ (listed in Table~2). Here, $L_{\rm FIR}$ is the
bolometric infrared luminosity integrated between $40-120~\mu$m. There is no
significant differences in the values of $q_{\rm FIR}$ and the slope when
compared with that of monochromatic 70$\mu$m. This is not surprising because,
as per the traditional definition of \citet{helou85}, the factor $3.75\times
10^{12}$ W (in Equation~\ref{def-q}) normalizes the integrated luminosity
between $40-120~\mu$m to the mean luminosity at $\sim80~\mu$m. This value is
close to 70$\mu$m.

\subsection{Detected red galaxies}

To assess the effect of including the 26 red galaxies on the radio--FIR
correlation, we have separately studied the red and blue galaxies that are
directly detected.  The two populations are well mixed in the radio--FIR
correlation and there are no systematic difference between them.  In Table~2 we
list the slope and the `$q$' parameter of the two classes estimated for
70$\mu$m, TIR and FIR. Both blue and red galaxies follow the correlation with
similar `$q$' parameters for all the three cases (i.e., $q_{\rm 70\mu m}$,
$q_{\rm TIR}$ and $q_{\rm FIR}$). However, the slope of the correlation is
found to be steeper in red galaxies as compared to the blue galaxies when
studied between $L_{\rm 1.4GHz}$ in the radio and $L_{\rm 70\mu m}$ and $L_{\rm
FIR}$ in the infrared (see Table~2). This difference is at less than $2\sigma$
significance and is perhaps caused due to relatively smaller sample size of the
red galaxies.  We do not observe any difference in the slope for blue and red
galaxies when studied for $L_{\rm TIR}$. Overall, including the red galaxies
in the sample of detected galaxies does not bias our results.

\begin{figure*}
\begin{center}
\begin{tabular}{cc}
{\mbox{\includegraphics[height=9cm]{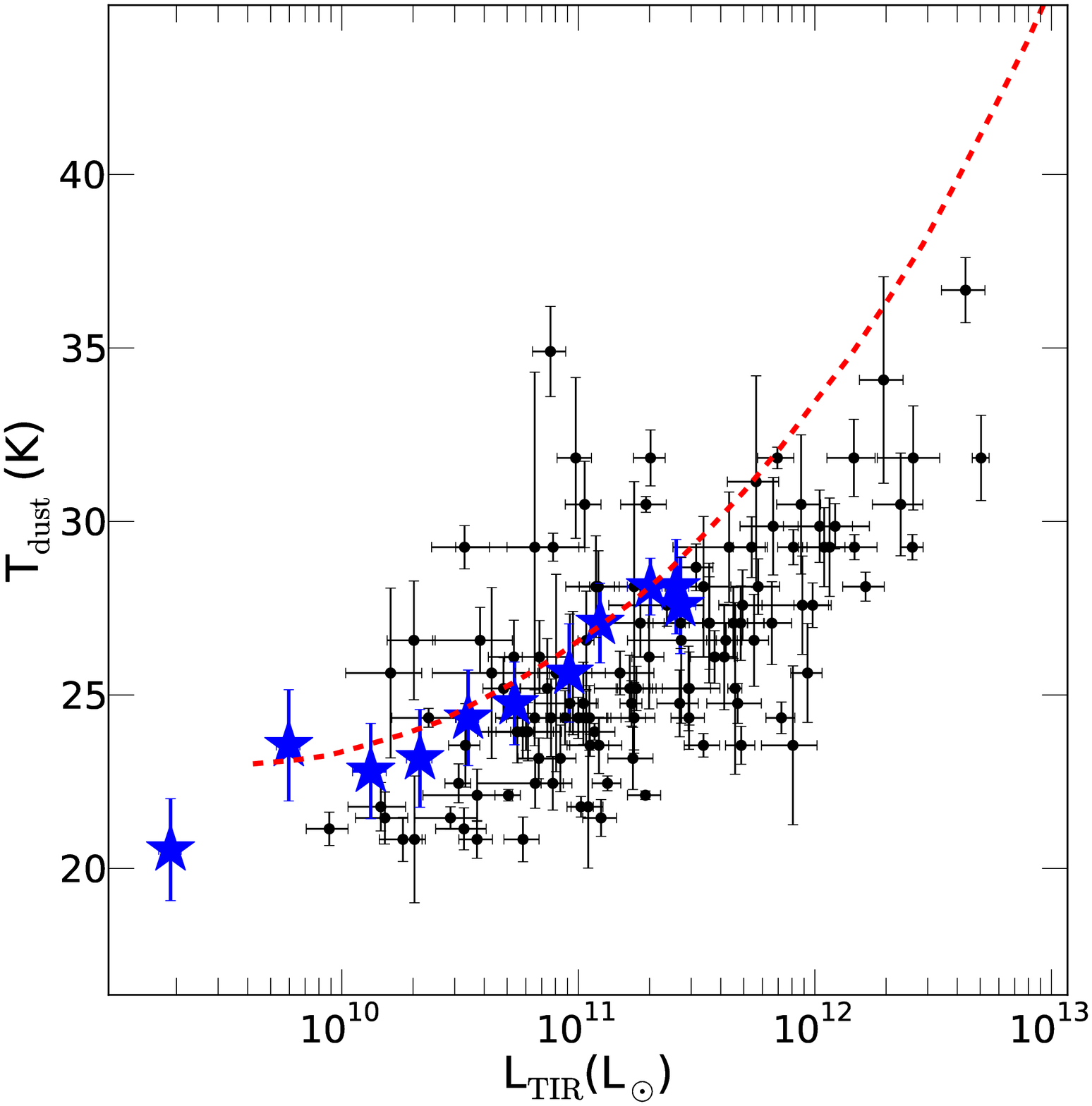}}}&
{\mbox{\includegraphics[height=9cm]{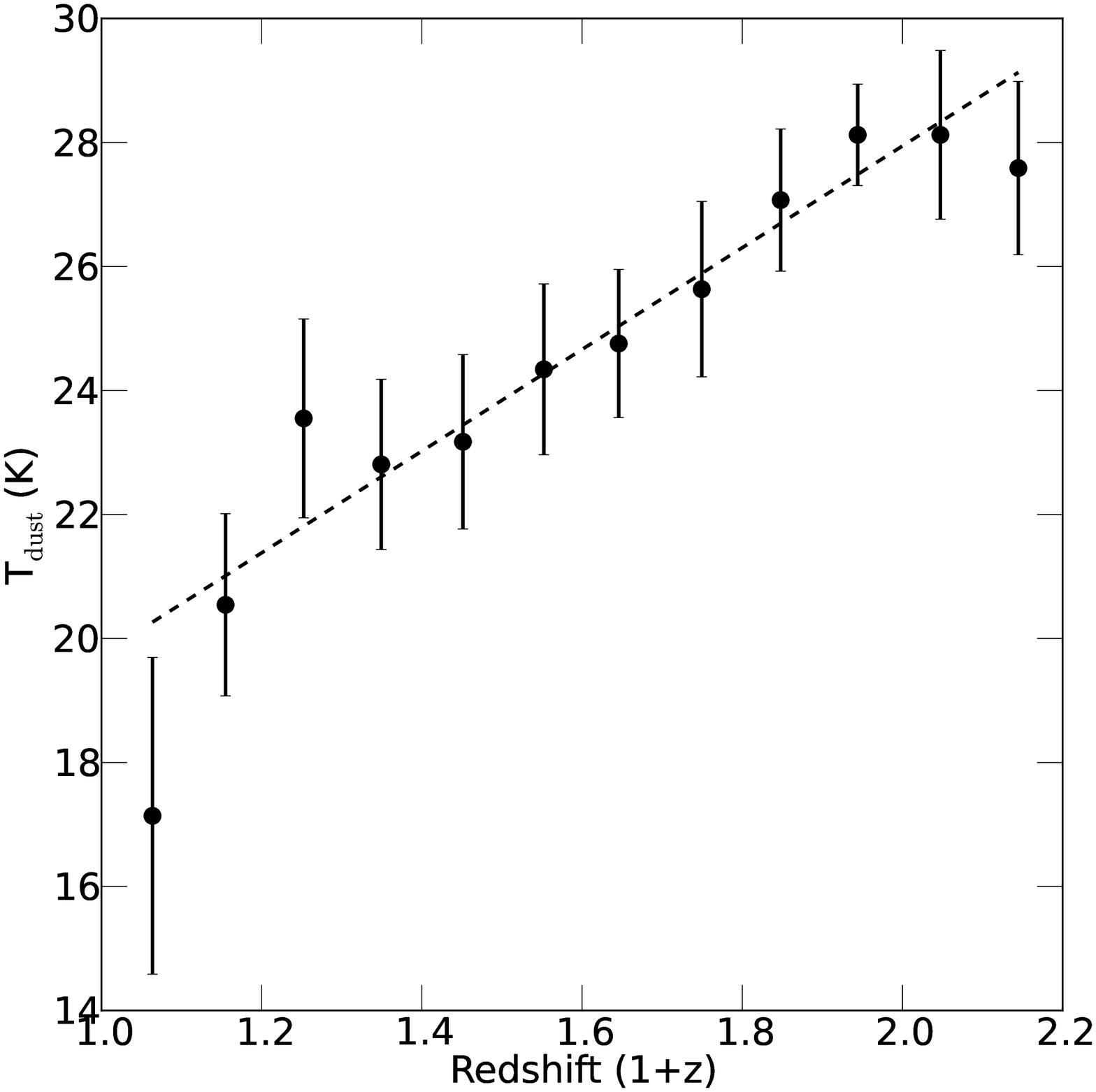}}}\\
\end{tabular}
\end{center}
\caption{{\it Left-hand panel} Variation of dust temperature with the total
infrared luminosity ($L_{\rm TIR}$) computed by integrating the SEDs between
8--1000 $\mu$m.  The blue stars represents stacked sources and circular points
are for detected sources.  The red dashed line shows the second order
polynomial fit between $0.2 < z < 0.5$ from \citet{magne14}. {\it Right hand
panel:} Variation of the estimated dust temperature ($T_{\rm dust}$) with
$(1+z)$. The dashed line shows the linear best fit of the form $T_{\rm dust} =
(8.2\pm0.9)(1+z) + (11.5\pm1.5)$.}
\label{tdust-tir}
\end{figure*}

Note that in our stacking analysis we have not distinguished between the
objects that have been detected as a PRIMUS counterpart or not. This is crucial
in estimating the typical mean flux or luminosity of the sample in each
redshift bin and waveband. Since, we have less than 50 percent detections w.r.t
PRIMUS objects, the median stacking could be insensitive to the objects having
PRIMUS counterparts. In our analysis, we find that the mean quantities are at an
average 20 percent higher than that of the median quantities. Our method of
stacking is equivalent to the quantity measured in Equation 3 of
\citet{magne14}. The difference between mean and median stacking is well
captured in our estimated bootstrap errors. The use of median luminosities does
not significantly affect the slope of the radio--FIR correlation shown in
Figure~\ref{frc} (left hand panel).  However, it can systematically increase
the value of $q_{\rm 70\mu m}$ and $q_{\rm TIR}$ of the stacks (the stars shown
on Figure~\ref{q}) by $\sim0.1$, i.e., $\sim5$ percent. This change in $q_{\rm
70\mu m}$ and $q_{\rm TIR}$ is well within the errors.

\section{Discussion}

We have studied the radio--FIR correlation of an optical flux-limited sample of
``blue cloud'' galaxies ($L_{\rm TIR}\sim 10^{9}-10^{11} L_\odot$) in the
redshift range $0.1-1.2$, using image stacking technique in luminosity space,
and for comparison, for luminous galaxies  ($L_{\rm TIR}\gtrsim 10^{11} L_\odot$) 
detected in the XMM-LSS field using deep GMRT observations at 0.325
GHz.  \citet{bourn11}, using a coarser redshift bin-size of $>0.2$ for
stacking, probed a similar luminosity regime for stellar-mass-selected
galaxies. We measure the mean luminosity of blue galaxies to be at least 2
orders of magnitude fainter than the galaxies that are directly detected in
radio and infrared in the XMM-LSS field and about three orders of magnitude
fainter than that of the previous studies \citep[see e.g.,][]{garre02, apple04,
kovac06, mao11}.  Our study probes a relatively poorly observed regime of the
radio--FIR correlation in luminosity and in redshift.

\subsection{Variation of dust temperature}

In Figure~\ref{tdust-tir} (left-hand panel), we show the variation of $T_{\rm
dust}$ with the total infrared luminosity ($L_{\rm TIR}$ in units of
$L_{\odot}$) for the stacked sources (shown as blue stars) and the detected
sources (shown as black points).  Our result is consistent with the second
order polynomial fit (shown as red dashed line) for slightly more luminous
galaxies between $0.2<z<0.5$ from \citet{magne14}. Note that the
\citet{magne14} curve does not extend to the faintest data point in our sample.

In Figure~\ref{tdust-tir} (right-hand panel) we plot the variation of the
characteristic dust temperature, $T_{\rm dust}$ with redshift, $1+z$. We find
$T_{\rm dust}$ to vary between $18-28$ K with $T_{\rm dust}$ increasing
linearly with redshift as,
\begin{equation}
	T_{\rm dust} = (8.2\pm0.9)(1+z) + (11.5\pm1.5)
\end{equation}
A similar linear variation of $T_{\rm dust}$ with redshift was reported by
\citet{kovac06} for distant sub-millimeter selected galaxies which are
$\gtrsim4$ orders of magnitude more luminous than our galaxies. However, such a
trend could be caused due to $T_{\rm dust}$ being correlated with $L_{\rm TIR}$
which in turn is correlated with redshift.

\subsection{Variation of `q'}

The mean value of $q_{\rm TIR}$ for the stacked sources is found to decrease
with redshift, and can be modeled as $q_{\rm TIR} \propto (1+z)^\gamma$
\citep[see, e.g.,][]{bourn11, iviso10a}, where $\gamma = -0.16\pm0.03$
(Figure~\ref{q} right-hand panel). The fitted value of $\gamma$ is similar to
$\gamma=-0.18\pm0.10$ found by \citet{bourn11} for galaxies with mass-limit
$\log(M_\star) > 10.5$ and $\gamma = -0.15\pm0.03$ by \citet{iviso10a} for
slightly luminous galaxies ($L_{\rm 1.4GHz}\sim10^{22}-10^{25}$ W Hz$^{-1}$).
This slight decrease in $q_{\rm TIR}$ with redshift, however, is not very
significant within the large errors. The correlation coefficient was estimated
taking into account the errors of $q_{\rm TIR}$, using a Monte Carlo method
wherein 1000 random samples of $q_{\rm TIR}$ were generated within the 1$\sigma$
error for each redshift bin. The values of $q_{\rm TIR}$ were then randomly
sampled from each redshift bin 1000 times to mimic different realizations and
the Spearman's rank correlation was estimated for each. We found the mean rank
correlation coefficient to be $\sim-0.13$. 

We do not observe such a trend for $q_{\rm 70\mu m}$.  This could be caused due
to the rest-frame luminosity at 70$\mu$m being dust temperature dependent.
Figure~\ref{tdust-tir} (right hand panel) shows an increase in dust temperature
with redshift resulting in increasing $L_{\rm 70\mu m}$, hence compensating for
any decrease in $q_{\rm 70\mu m}$.  This trend is consistent with recent
observations of \citet{smith14}, where they find monochromatic $q$ near the
peak of the dust emission to remain constant with dust temperature.

\subsection{Is `$q$' an indicator of evolution of the correlation?}

In our study, we find the radio--FIR correlation to hold true within the
redshift range 0.1--1.2 for blue galaxies and also for much luminous galaxies
spanning about 5 orders of magnitude. The correlation can be fitted by a single
slope, $b = 1.09\pm0.05$, when studied with monochromatic FIR luminosity
($L_{\rm 70\mu m}$). For bolometric FIR luminosity ($L_{\rm TIR}$), we find,
$b=1.11\pm0.04$ (see Table~2).  For our sample, we find that the slope of
the radio--FIR correlation is systematically steeper than unity with
1.8$\sigma$ significance for $L_{\rm 70\mu m}$ and 2.75$\sigma$ for
$L_{\rm TIR}$.

The non-linear slope of the radio--FIR correlation brings to light the
ambiguity in use of the quantity `$q$' (both $q_{\rm 70\mu m}$ and $q_{\rm
TIR}$) in quantifying the tightness of the correlation or in studying evolution
of the radio--FIR correlation. Note that, apart from the traditional
definition, $q$ can also be written as, 
\begin{equation}
q = -\left(\frac{1}{b}\right)\log_{10}a + \left(\frac{1-b}{b}\right) \log_{10}L_{\rm 1.4GHz}
\label{qaltdef}
\end{equation}
Here, $a$ is the proportionality constant of the radio--FIR correlation, i.e.,
$L_{\rm 1.4GHz} = a L_{\rm IR}^b$. This relation holds for both monochromatic
$q_{\rm 70\mu m}$ and bolometric $q_{\rm TIR}$. Clearly, for non-unity slopes,
the absolute value of $q$ would depend on the slope and also on radio
luminosity and hence cannot be assumed to be a constant. Several other works in
the literature have indeed reported nonlinear slopes \citep[][]{price92,
nikla97b, bell03}.  Moreover, $L_{\rm 1.4GHz}$ in the rest-frame depends on the
assumed value of the spectral index and thus can affect the value of $q$
further. It is, therefore, difficult to compare the value of the parameter $q$
in the literature and interpret any variation of $q$ as evolution of the
correlation. This non-linearity of the slope could also give rise to the
observed slight change in $q_{\rm TIR}$ (see Section~4 and Figure~\ref{q}). We
have modeled the variation as $q_{\rm TIR} = 2.53~(1+z)^{-0.16}$, which
results in $\Delta q_{\rm TIR} \approx -0.26$ between $z =$ 0 and 1.  This
change has been observed in other studies, e.g., \citet{iviso10a, bourn11,
magne12}.  Following Equation~\ref{qaltdef}, the expected change in $q_{\rm
TIR}$ is given by, 
\begin{equation}
	\Delta q_{\rm TIR} = \left(\frac{1-b}{b}\right)
\log_{10}\left(\frac{L_{\rm 1.4GHz}\mid_{z=1}}{L_{\rm 1.4 GHz}\mid_{z=0}}\right).
\end{equation}
Here, $L_{\rm 1.4 GHz}\mid_{z}$ is the rest-frame 1.4 GHz luminosity at a
redshift $z$.  From Figure~\ref{frc} we find $L_{\rm 1.4GHz}$ to increase by a
factor $\sim 10^3$ between $z=$ 0 and 1 for the stacked galaxies.  Thus,
$\Delta q_{\rm TIR}\approx -0.27$ for $b=1.1$ (see Table~2), similar to the
observed change. It is therefore difficult to interpret any change in the
parameter `$q$' as an evolution of the radio--FIR correlation for non-linear
slopes. Note that, although our results are based on a sample that
is flux limited and color selected, it is important to ascertain the 
slope for a more representative sample to conclude any evolution based on
`$q$'.

The parameter `$q$', explicitly depends on the absolute values of the ISM
parameters, such as, density of dust ($\rm \rho_{\rm dust}$), $T_{\rm dust}$,
dust emissivity ($Q$), properties of dust grains, like size ($a$), CRE density
($n_{\rm CRe}$), magnetic field ($B$) and the non-thermal spectral index
($\alpha_{\rm nt}$). In terms of these quantities $q$ can be written as;
\begin{equation}
	q \sim \frac{\rho_{\rm dust}Q(\lambda, a) B_\lambda(T_{\rm dust})}{n_{\rm CRe} B^{1+\alpha_{\rm nt}}}
\end{equation}
Hence, based on a controlled sample `$q$' can be used as a proxy to
`relatively' distinguish object based on their ISM properties.

On the other hand, the slope, $b$, is an important parameter that connects the
various physical parameters in the ISM \citep[see e.g.,][]{nikla97b, dumas11,
schle13}. Note that at the heart of the radio--FIR correlation lies the
interdependence between various ISM parameters and not their absolute values,
{\it viz.}, coupling between $B$ and gas density \citep[$B\propto \rho_{\rm
gas}^\kappa$; see e.g.,][]{helou93, nikla97b, grove03}, the Kennicutt-Schmidt
law \citep{kenni98} and spectrum of the CREs.  Therefore, the evolution in
slope can throw meaningful insights into understanding the cause and/or
evolution of the radio--FIR correlation.  In our study, a single slope is
enough to fit the radio--FIR correlation throughout the luminosity and redshift
range.  We do not find any evidence of its evolution within the uncertainty of
our measurements. 

\subsection{Heterogeneity in the sample}

In spite of the wealth of observational evidence, a clear understanding of the
reason behind the radio--FIR correlation, across various galaxy types with wide
range of star-formation activity, magnetic field strengths, nature of ISM
turbulence, etc., remains elusive. One of the major areas of progress in recent
years has been in our understanding of how star-formation/supernova driven
turbulence helps in amplification of magnetic field strength in galaxies
\citep[][]{breit09, deavi05,maclo05, gent13a, gent13b}, and how the magnetic
field in turn couples with the gas density, and hence the star formation rate
\citep{cho00, grove03}. Recently, \citet{schle13} interpreted the radio--FIR
correlation as a result of turbulent magnetic field amplification driven by
star-formation activity and their interplay. Cosmic ray energy loss mechanisms
and magnetic field strength evolution at higher redshifts may affect the
synchrotron emission and hence change the non-thermal spectral index giving rise
to modification of the radio--FIR correlation.  

We would like to point out, since our sample is flux-limited in the
optical, we have a wide range of galaxies in our sample. In terms of $L_{\rm
TIR}$, galaxies are classified as, 1) normal star-forming ($L_{\rm TIR}
\lesssim 10^{11} L_\odot$), 2) LIRGs ($10^{11}L_\odot < L_{\rm TIR} <
10^{12}$), 3) ULIRGs ($10^{12}L_\odot < L_{\rm TIR} < 10^{13}$) and 4) Hyper
LIRGs ($L_{\rm TIR} > 10^{13}$) \citep[see e.g.,][]{sande96}. As per these
definitions, the stacked sources mostly fall under the category of normal
star-forming galaxies.  Although, the number of (U)LIRGs are significantly less
in the local universe ($z < 0.3$), they could contribute significantly to the
infrared luminosity at higher redshifts \citep{kim98, caput07, eser14}.  
\citet{magne09} found significant evolution of the contribution from (U)LIRGs
to the cosmic star formation rate density with respect to that from main
sequence galaxies. Thus, towards the higher redshift bins ($z\gtrsim0.9$)
there could be significant contribution from LIRGs and ULIRGs to the stacked
luminosity.  On the other hand, the majority of directly detected sources falls
under the LIRG and ULIRG category.

In this study, the stacked blue galaxies and the directly detected luminous
galaxies are expected to have different nature of ISM turbulence and magnetic
field amplification.  For normal star-forming galaxies, at lower luminosity
range, the turbulence is mostly driven by star-formation activity while for the
luminous galaxies the turbulence could be driven by interactions and mergers
\citep{veill02, eser14}.  Even perhaps, the mechanisms driving the radio--FIR
correlation can differ in these wide variety of galaxies
\citep[see][]{lacki10}.  Ideally, one should study the different population of
galaxies based on a well defined luminosity or stellar-mass selection.
Unfortunately, due to the lack of sufficient number of directly detected
galaxies of similar luminosities/stellar-masses across all redshift bins, our
sample does not allow us to study this aspect.  But, nonetheless, notable fact
is that both class of galaxies follow the radio--FIR correlation with similar
parameters.  This brings to light the universal nature of the radio--FIR
correlation and suggests similar global processes at play in galaxies with very
different ISM properties.

The stacking technique described in this paper trades off sample size for
improved signal to noise. Such a compromise has the obvious drawback that one
can only study the mean properties of a sample. Outliers in the parameter space
that correspond to more extreme physical environments are completely missed
out.

In order to study the radio-FIR correlation in normal galaxies at high redshift
without the above trade-off requires large, spectroscopically confirmed samples
of such galaxies spanning the entire redshift range of interest. Multi-object
spectrographs on large optical telescopes are now routinely producing such
samples totalling hundreds of thousands of galaxies (e.g., \citealt{coil11}
[PRIMUS]; \citealt{guzzo13} [VIPERS]; \citealt{newma13} [DEEP2];
\citealt{lefev05} [VVDS]; \citealt{baldr10} [GAMA]). Radio followup at
sufficient depth will also be routine with the JVLA \citep{jarvi14} and
deep surveys with upcoming facilities such as ASKAP \citep{norri11} and
MeerKAT. The situation with deep FIR surveys is less hopeful. The Level 5 and
Level 6 HerMES data are already sufficiently confused that increased depth in
the FIR can only be achieved with a larger telescope operating at somewhat
shorter wavelengths. Such a super-Spitzer will need to have a large aperture
and instruments far more sensitive than MIPS.  Unfortunately, such a telescope
is nowhere on the horizon.

The stacking methods described in this paper can be gainfully applied to a
number of other deep fields where almost all the required data are available in
public archives.

\acknowledgements  

We thank the anonymous referee for insightful comments that greatly improved
the content and presentation of this paper.  We thank Nissim Kanekar for
fruitful suggestions on sample selection. We thank Dominik Schleicher, Rainer
Beck, Nirupam Roy and Dipanjan Mitra for useful discussions.  We gratefully
acknowledge generous support from the Indo-French Center for the Promotion of
Advanced Research (Centre Franco- Indien pour la Promotion de la Recherche
Avance) under program no. 4404- 3. We thank Chris Simpson for providing the
1.4-GHz VLA radio image of the SXDF field. This work has made use of the PRIMUS
for which the funding has been provided by NSF grants AST-0607701, 0908246,
0908442, 0908354, and NASA grant 08-ADP08-0019.

%\bibliographystyle{apj}
%\bibliography{references}

\begin{thebibliography}{99}

\bibitem[{{Appleton} {et~al.}(2004){Appleton}, {Fadda}, {Marleau}, {Frayer},
  {Helou}, {Condon}, {Choi}, {Yan}, {Lacy}, {Wilson}, {Armus}, {Chapman},
  {Fang}, {Heinrichson}, {Im}, {Jannuzi}, {Storrie-Lombardi}, {Shupe},
  {Soifer}, {Squires}, \& {Teplitz}}]{apple04}
{Appleton}, P.~N., {et~al.} 2004, \apjs, 154, 147

\bibitem[{{Baldry} {et~al.}(2006){Baldry}, {Balogh}, {Bower}, {Glazebrook},
  {Nichol}, {Bamford}, \& {Budavari}}]{baldr06}
{Baldry}, I.~K., {Balogh}, M.~L., {Bower}, R.~G., {Glazebrook}, K., {Nichol},
  R.~C., {Bamford}, S.~P., \& {Budavari}, T. 2006, \mnras, 373, 469

\bibitem[{{Baldry} {et~al.}(2010){Baldry}, {Robotham}, {Hill}, {Driver},
  {Liske}, {Norberg}, {Bamford}, {Hopkins}, {Loveday}, {Peacock}, {Cameron},
  {Croom}, {Cross}, {Doyle}, {Dye}, {Frenk}, {Jones}, {van Kampen}, {Kelvin},
  {Nichol}, {Parkinson}, {Popescu}, {Prescott}, {Sharp}, {Sutherland},
  {Thomas}, \& {Tuffs}}]{baldr10}
{Baldry}, I.~K., {et~al.} 2010, \mnras, 404, 86

\bibitem[{{Basu} {et~al.}(2012{\natexlab{a}}){Basu}, {Mitra}, {Wadadekar}, \&
  {Ishwara-Chandra}}]{basu12a}
{Basu}, A., {Mitra}, D., {Wadadekar}, Y., \& {Ishwara-Chandra}, C.~H.
  2012{\natexlab{a}}, \mnras, 419, 1136

\bibitem[{{Basu} {et~al.}(2012{\natexlab{b}}){Basu}, {Roy}, \&
  {Mitra}}]{basu12b}
{Basu}, A., {Roy}, S., \& {Mitra}, D. 2012{\natexlab{b}}, \apj, 756, 141

\bibitem[{{Bell}(2003)}]{bell03}
{Bell}, E.~F. 2003, \apj, 586, 794

\bibitem[{{B{\'e}thermin} {et~al.}(2010){B{\'e}thermin}, {Dole}, {Cousin}, \&
  {Bavouzet}}]{bethe10}
{B{\'e}thermin}, M., {Dole}, H., {Cousin}, M., \& {Bavouzet}, N. 2010, \aap,
  516, A43

\bibitem[{{B{\'e}thermin} {et~al.}(2012){B{\'e}thermin}, {Le Floc'h}, {Ilbert},
  {Conley}, {Lagache}, {Amblard}, {Arumugam}, {Aussel}, {Berta}, {Bock},
  {Boselli}, {Buat}, {Casey}, {Castro-Rodr{\'{\i}}guez}, {Cava}, {Clements},
  {Cooray}, {Dowell}, {Eales}, {Farrah}, {Franceschini}, {Glenn}, {Griffin},
  {Hatziminaoglou}, {Heinis}, {Ibar}, {Ivison}, {Kartaltepe}, {Levenson},
  {Magdis}, {Marchetti}, {Marsden}, {Nguyen}, {O'Halloran}, {Oliver}, {Omont},
  {Page}, {Panuzzo}, {Papageorgiou}, {Pearson}, {P{\'e}rez-Fournon}, {Pohlen},
  {Rigopoulou}, {Roseboom}, {Rowan-Robinson}, {Salvato}, {Schulz}, {Scott},
  {Seymour}, {Shupe}, {Smith}, {Symeonidis}, {Trichas}, {Tugwell}, {Vaccari},
  {Valtchanov}, {Vieira}, {Viero}, {Wang}, {Xu}, \& {Zemcov}}]{bethe12}
{B{\'e}thermin}, M., {et~al.} 2012, \aap, 542, A58

\bibitem[{{Bondi} {et~al.}(2003){Bondi}, {Ciliegi}, {Zamorani}, {Gregorini},
  {Vettolani}, {Parma}, {de Ruiter}, {Le Fevre}, {Arnaboldi}, {Guzzo},
  {Maccagni}, {Scaramella}, {Adami}, {Bardelli}, {Bolzonella}, {Bottini},
  {Cappi}, {Foucaud}, {Franzetti}, {Garilli}, {Gwyn}, {Ilbert}, {Iovino}, {Le
  Brun}, {Marano}, {Marinoni}, {McCracken}, {Meneux}, {Pollo}, {Pozzetti},
  {Radovich}, {Ripepi}, {Rizzo}, {Scodeggio}, {Tresse}, {Zanichelli}, \&
  {Zucca}}]{bondi03}
{Bondi}, M., {et~al.} 2003, \aap, 403, 857

\bibitem[{{Bourne} {et~al.}(2011){Bourne}, {Dunne}, {Ivison}, {Maddox},
  {Dickinson}, \& {Frayer}}]{bourn11}
{Bourne}, N., {Dunne}, L., {Ivison}, R.~J., {Maddox}, S.~J., {Dickinson}, M.,
  \& {Frayer}, D.~T. 2011, \mnras, 410, 1155

\bibitem[{{Brammer} {et~al.}(2011){Brammer}, {Whitaker}, {van Dokkum},
  {Marchesini}, {Franx}, {Kriek}, {Labb{\'e}}, {Lee}, {Muzzin}, {Quadri},
  {Rudnick}, \& {Williams}}]{bramm11}
{Brammer}, G.~B., {et~al.} 2011, \apj, 739, 24

\bibitem[{{Breitschwerdt} {et~al.}(2009){Breitschwerdt}, {de Avillez}, {Fuchs},
  \& {Dettbarn}}]{breit09}
{Breitschwerdt}, D., {de Avillez}, M.~A., {Fuchs}, B., \& {Dettbarn}, C. 2009,
  \ssr, 143, 263

\bibitem[{{Caputi} {et~al.}(2007){Caputi}, {Lagache}, {Yan}, {Dole},
  {Bavouzet}, {Le Floc'h}, {Choi}, {Helou}, \& {Reddy}}]{caput07}
{Caputi}, K.~I., {et~al.} 2007, \apj, 660, 97

\bibitem[{{Casey}(2012)}]{casey12}
{Casey}, C.~M. 2012, \mnras, 425, 3094

\bibitem[{{Cho} \& {Vishniac}(2000)}]{cho00}
{Cho}, J., \& {Vishniac}, E.~T. 2000, \apj, 539, 273

\bibitem[{{Chy{\.z}y} {et~al.}(2011){Chy{\.z}y}, {We{\.z}gowiec}, {Beck}, \&
  {Bomans}}]{chyzy11}
{Chy{\.z}y}, K.~T., {We{\.z}gowiec}, M., {Beck}, R., \& {Bomans}, D.~J. 2011,
  \aap, 529, A94

\bibitem[{{Coil} {et~al.}(2011){Coil}, {Blanton}, {Burles}, {Cool},
  {Eisenstein}, {Moustakas}, {Wong}, {Zhu}, {Aird}, {Bernstein}, {Bolton}, \&
  {Hogg}}]{coil11}
{Coil}, A.~L., {et~al.} 2011, \apj, 741, 8

\bibitem[{{Condon}(1992)}]{condo92}
{Condon}, J.~J. 1992, \araa, 30, 575

\bibitem[{{Condon} {et~al.}(1991){Condon}, {Huang}, {Yin}, \&
  {Thuan}}]{condo91b}
{Condon}, J.~J., {Huang}, Z.-P., {Yin}, Q.~F., \& {Thuan}, T.~X. 1991, \apj,
  378, 65

\bibitem[{{Cool} {et~al.}(2013){Cool}, {Moustakas}, {Blanton}, {Burles},
  {Coil}, {Eisenstein}, {Wong}, {Zhu}, {Aird}, {Bernstein}, {Bolton}, {Hogg},
  \& {Mendez}}]{cool13}
{Cool}, R.~J., {et~al.} 2013, \apj, 767, 118

\bibitem[{{de Avillez} \& {Breitschwerdt}(2005)}]{deavi05}
{de Avillez}, M.~A., \& {Breitschwerdt}, D. 2005, \aap, 436, 585

\bibitem[{{Del Moro} {et~al.}(2013){Del Moro}, {Alexander}, {Mullaney},
  {Daddi}, {Pannella}, {Bauer}, {Pope}, {Dickinson}, \& et~al.}]{delmo13}
{Del Moro}, A., {et~al.} 2013, \aap, 549, A59

\bibitem[{{Dumas} {et~al.}(2011){Dumas}, {Schinnerer}, {Tabatabaei}, {Beck},
  {Velusamy}, \& {Murphy}}]{dumas11}
{Dumas}, G., {Schinnerer}, E., {Tabatabaei}, F.~S., {Beck}, R., {Velusamy}, T.,
  \& {Murphy}, E. 2011, \aj, 141, 41

\bibitem[{Efron \& Tibshirani(1994)}]{efron94}
Efron, B., \& Tibshirani, R. 1994, An Introduction to the Bootstrap, Chapman \&
  Hall/CRC Monographs on Statistics \& Applied Probability (Taylor \& Francis)

\bibitem[{{Garrett}(2002)}]{garre02}
{Garrett}, M.~A. 2002, \aap, 384, L19

\bibitem[{{Gent} {et~al.}(2013{\natexlab{a}}){Gent}, {Shukurov}, {Fletcher},
  {Sarson}, \& {Mantere}}]{gent13a}
{Gent}, F.~A., {Shukurov}, A., {Fletcher}, A., {Sarson}, G.~R., \& {Mantere},
  M.~J. 2013{\natexlab{a}}, \mnras, 432, 1396

\bibitem[{{Gent} {et~al.}(2013{\natexlab{b}}){Gent}, {Shukurov}, {Sarson},
  {Fletcher}, \& {Mantere}}]{gent13b}
{Gent}, F.~A., {Shukurov}, A., {Sarson}, G.~R., {Fletcher}, A., \& {Mantere},
  M.~J. 2013{\natexlab{b}}, \mnras, 430, L40

\bibitem[{{Groves} {et~al.}(2003){Groves}, {Cho}, {Dopita}, \&
  {Lazarian}}]{grove03}
{Groves}, B.~A., {Cho}, J., {Dopita}, M., \& {Lazarian}, A. 2003, PASA, 20,
  252

\bibitem[{{Guzzo} {et~al.}(2013){Guzzo}, {Scodeggio}, {Garilli}, {Granett},
  {Abbas}, {Adami}, {Arnouts}, {Bel}, \& et~al.}]{guzzo13}
{Guzzo}, L., {et~al.} 2013, ArXiv e-prints

\bibitem[{{Helou} \& {Bicay}(1993)}]{helou93}
{Helou}, G., \& {Bicay}, M.~D. 1993, \apj, 415, 93

\bibitem[{{Helou} {et~al.}(1985){Helou}, {Soifer}, \&
  {Rowan-Robinson}}]{helou85}
{Helou}, G., {Soifer}, B.~T., \& {Rowan-Robinson}, M. 1985, \apjl, 298, L7

\bibitem[{{Hodge} {et~al.}(2011){Hodge}, {Becker}, {White}, {Richards}, \&
  {Zeimann}}]{hodge11}
{Hodge}, J.~A., {Becker}, R.~H., {White}, R.~L., {Richards}, G.~T., \&
  {Zeimann}, G.~R. 2011, \aj, 142, 3

\bibitem[{{Hoernes} {et~al.}(1998){Hoernes}, {Berkhuijsen}, \& {Xu}}]{hoern98}
{Hoernes}, P., {Berkhuijsen}, E.~M., \& {Xu}, C. 1998, \aap, 334, 57

\bibitem[{{Hughes} {et~al.}(2006){Hughes}, {Wong}, {Ekers}, {Staveley-Smith},
  {Filipovic}, {Maddison}, {Fukui}, \& {Mizuno}}]{hughe06}
{Hughes}, A., {Wong}, T., {Ekers}, R., {Staveley-Smith}, L., {Filipovic}, M.,
  {Maddison}, S., {Fukui}, Y., \& {Mizuno}, N. 2006, \mnras, 370, 363

\bibitem[{{Ilbert} {et~al.}(2006){Ilbert}, {Arnouts}, {McCracken},
  {Bolzonella}, {Bertin}, {Le F{\`e}vre}, {Mellier}, {Zamorani}, {Pell{\`o}},
  {Iovino}, {Tresse}, {Le Brun}, {Bottini}, {Garilli}, {Maccagni}, {Picat},
  {Scaramella}, {Scodeggio}, {Vettolani}, {Zanichelli}, {Adami}, {Bardelli},
  {Cappi}, {Charlot}, {Ciliegi}, {Contini}, {Cucciati}, {Foucaud}, {Franzetti},
  {Gavignaud}, {Guzzo}, {Marano}, {Marinoni}, {Mazure}, {Meneux}, {Merighi},
  {Paltani}, {Pollo}, {Pozzetti}, {Radovich}, {Zucca}, {Bondi}, {Bongiorno},
  {Busarello}, {de La Torre}, {Gregorini}, {Lamareille}, {Mathez}, {Merluzzi},
  {Ripepi}, {Rizzo}, \& {Vergani}}]{ilber06}
{Ilbert}, O., {et~al.} 2006, \aap, 457, 841

\bibitem[{{Isobe} {et~al.}(1990){Isobe}, {Feigelson}, {Akritas}, \&
  {Babu}}]{isobe90}
{Isobe}, T., {Feigelson}, E.~D., {Akritas}, M.~G., \& {Babu}, G.~J. 1990, \apj,
  364, 104

\bibitem[{{Ivison} {et~al.}(2010{\natexlab{a}}){Ivison}, {Alexander}, {Biggs},
  {Brandt}, {Chapin}, {Coppin}, {Devlin}, {Dickinson}, {Dunlop}, {Dye},
  {Eales}, {Frayer}, {Halpern}, {Hughes}, {Ibar}, {Kov{\'a}cs}, {Marsden},
  {Moncelsi}, {Netterfield}, {Pascale}, {Patanchon}, {Rafferty}, {Rex},
  {Schinnerer}, {Scott}, {Semisch}, {Smail}, {Swinbank}, {Truch}, {Tucker},
  {Viero}, {Walter}, {Wei{\ss}}, {Wiebe}, \& {Xue}}]{iviso10a}
{Ivison}, R.~J., {et~al.} 2010{\natexlab{a}}, \mnras, 402, 245

\bibitem[{{Ivison} {et~al.}(2010{\natexlab{b}}){Ivison}, {Magnelli}, {Ibar},
  {Andreani}, {Elbaz}, {Altieri}, {Amblard}, {Arumugam}, {Auld}, {Aussel},
  {Babbedge}, {Berta}, {Blain}, {Bock}, {Bongiovanni}, {Boselli}, {Buat},
  {Burgarella}, {Castro-Rodr{\'{\i}}guez}, {Cava}, {Cepa}, {Chanial},
  {Cimatti}, {Cirasuolo}, {Clements}, {Conley}, {Conversi}, {Cooray}, {Daddi},
  {Dominguez}, {Dowell}, {Dwek}, {Eales}, {Farrah}, {F{\"o}rster Schreiber},
  {Fox}, {Franceschini}, {Gear}, {Genzel}, {Glenn}, {Griffin}, {Gruppioni},
  {Halpern}, {Hatziminaoglou}, {Isaak}, {Lagache}, {Levenson}, {Lu}, {Lutz},
  {Madden}, {Maffei}, {Magdis}, {Mainetti}, {Maiolino}, {Marchetti},
  {Morrison}, {Mortier}, {Nguyen}, {Nordon}, {O'Halloran}, {Oliver}, {Omont},
  {Owen}, {Page}, {Panuzzo}, {Papageorgiou}, {Pearson}, {P{\'e}rez-Fournon},
  {P{\'e}rez Garc{\'{\i}}a}, {Poglitsch}, {Pohlen}, {Popesso}, {Pozzi},
  {Rawlings}, {Raymond}, {Rigopoulou}, {Riguccini}, {Rizzo}, {Rodighiero},
  {Roseboom}, {Rowan-Robinson}, {Saintonge}, {Sanchez Portal}, {Santini},
  {Schulz}, {Scott}, {Seymour}, {Shao}, {Shupe}, {Smith}, {Stevens}, {Sturm},
  {Symeonidis}, {Tacconi}, {Trichas}, {Tugwell}, {Vaccari}, {Valtchanov},
  {Vieira}, {Vigroux}, {Wang}, {Ward}, {Wright}, {Xu}, \& {Zemcov}}]{iviso10b}
---. 2010{\natexlab{b}}, \aap, 518, L31

\bibitem[{{Jarvis} {et~al.}(2014){Jarvis}, {Bhatnagar}, {Bruggen}, {Ferrari},
  {Heywood}, {Hardcastle}, {Murphy}, {Taylor}, {Smirnov}, {Simpson}, {Smolcic},
  {Stil}, \& {van der Heyden}}]{jarvi14}
{Jarvis}, M.~J., {et~al.} 2014, ArXiv e-prints:1401.4018

\bibitem[{{Karim} {et~al.}(2011){Karim}, {Schinnerer},
  {Mart{\'{\i}}nez-Sansigre}, {Sargent}, {van der Wel}, {Rix}, {Ilbert},
  {Smol{\v c}i{\'c}}, {Carilli}, {Pannella}, {Koekemoer}, {Bell}, \&
  {Salvato}}]{karim11}
{Karim}, A., {et~al.} 2011, \apj, 730, 61

\bibitem[{{Kauffmann} {et~al.}(2003){Kauffmann}, {Heckman}, {White}, {Charlot},
  {Tremonti}, {Peng}, {Seibert}, {Brinkmann}, {Nichol}, {SubbaRao}, \&
  {York}}]{kauff03}
{Kauffmann}, G., {et~al.} 2003, \mnras, 341, 54

\bibitem[{{Kennicutt}(1998)}]{kenni98}
{Kennicutt}, Jr., R.~C. 1998, \araa, 36, 189

\bibitem[{{Kilerci Eser} {et~al.}(2014){Kilerci Eser}, {Goto}, \&
  {Doi}}]{eser14}
{Kilerci Eser}, E., {Goto}, T., \& {Doi}, Y. 2014, \apj, 797, 54

\bibitem[{{Kim} \& {Sanders}(1998)}]{kim98}
{Kim}, D.-C., \& {Sanders}, D.~B. 1998, \apjs, 119, 41

\bibitem[{{Kov{\'a}cs} {et~al.}(2006){Kov{\'a}cs}, {Chapman}, {Dowell},
  {Blain}, {Ivison}, {Smail}, \& {Phillips}}]{kovac06}
{Kov{\'a}cs}, A., {Chapman}, S.~C., {Dowell}, C.~D., {Blain}, A.~W., {Ivison},
  R.~J., {Smail}, I., \& {Phillips}, T.~G. 2006, \apj, 650, 592

\bibitem[{{Labb{\'e}} {et~al.}(2007){Labb{\'e}}, {Franx}, {Rudnick},
  {Schreiber}, {van Dokkum}, {Moorwood}, {Rix}, {R{\"o}ttgering}, {Trujillo},
  \& {van der Werf}}]{labbe07}
{Labb{\'e}}, I., {et~al.} 2007, \apj, 665, 944

\bibitem[{{Lacki} {et~al.}(2010){Lacki}, {Thompson}, \& {Quataert}}]{lacki10}
{Lacki}, B.~C., {Thompson}, T.~A., \& {Quataert}, E. 2010, \apj, 717, 1

\bibitem[{{Le F{\`e}vre} {et~al.}(2005){Le F{\`e}vre}, {Vettolani}, {Garilli},
  {Tresse}, {Bottini}, {Le Brun}, {Maccagni}, {Picat}, {Scaramella},
  {Scodeggio}, {Zanichelli}, {Adami}, {Arnaboldi}, {Arnouts}, {Bardelli},
  {Bolzonella}, {Cappi}, {Charlot}, {Ciliegi}, {Contini}, {Foucaud},
  {Franzetti}, {Gavignaud}, {Guzzo}, {Ilbert}, {Iovino}, {McCracken}, {Marano},
  {Marinoni}, {Mathez}, {Mazure}, {Meneux}, {Merighi}, {Paltani}, {Pell{\`o}},
  {Pollo}, {Pozzetti}, {Radovich}, {Zamorani}, {Zucca}, {Bondi}, {Bongiorno},
  {Busarello}, {Lamareille}, {Mellier}, {Merluzzi}, {Ripepi}, \&
  {Rizzo}}]{lefev05}
{Le F{\`e}vre}, O., {et~al.} 2005, \aap, 439, 845

\bibitem[{{Lonsdale} {et~al.}(2003){Lonsdale}, {Smith}, {Rowan-Robinson},
  {Surace}, {Shupe}, {Xu}, {Oliver}, {Padgett}, \& et~al.}]{lonsd03}
{Lonsdale}, C.~J., {et~al.} 2003, \pasp, 115, 897

\bibitem[{{Mac Low} {et~al.}(2005){Mac Low}, {Balsara}, {Kim}, \& {de
  Avillez}}]{maclo05}
{Mac Low}, M.-M., {Balsara}, D.~S., {Kim}, J., \& {de Avillez}, M.~A. 2005,
  \apj, 626, 864

\bibitem[{{Magnelli} {et~al.}(2009){Magnelli}, {Elbaz}, {Chary}, {Dickinson},
  {Le Borgne}, {Frayer}, \& {Willmer}}]{magne09}
{Magnelli}, B., {Elbaz}, D., {Chary}, R.~R., {Dickinson}, M., {Le Borgne}, D.,
  {Frayer}, D.~T., \& {Willmer}, C.~N.~A. 2009, \aap, 496, 57

\bibitem[{{Magnelli} {et~al.}(2012){Magnelli}, {Lutz}, {Santini}, {Saintonge},
  {Berta}, {Albrecht}, {Altieri}, {Andreani}, {Aussel}, {Bertoldi},
  {B{\'e}thermin}, {Bongiovanni}, {Capak}, {Chapman}, {Cepa}, {Cimatti},
  {Cooray}, {Daddi}, {Danielson}, {Dannerbauer}, {Dunlop}, {Elbaz}, {Farrah},
  {F{\"o}rster Schreiber}, {Genzel}, {Hwang}, {Ibar}, {Ivison}, {Le Floc'h},
  {Magdis}, {Maiolino}, {Nordon}, {Oliver}, {P{\'e}rez Garc{\'{\i}}a},
  {Poglitsch}, {Popesso}, {Pozzi}, {Riguccini}, {Rodighiero}, {Rosario},
  {Roseboom}, {Salvato}, {Sanchez-Portal}, {Scott}, {Smail}, {Sturm},
  {Swinbank}, {Tacconi}, {Valtchanov}, {Wang}, \& {Wuyts}}]{magne12}
{Magnelli}, B., {et~al.} 2012, \aap, 539, A155

\bibitem[{{Magnelli} {et~al.}(2014){Magnelli}, {Lutz}, {Saintonge}, {Berta},
  {Santini}, {Symeonidis}, {Altieri}, {Andreani}, {Aussel}, {B{\'e}thermin},
  {Bock}, {Bongiovanni}, {Cepa}, {Cimatti}, {Conley}, {Daddi}, {Elbaz},
  {F{\"o}rster Schreiber}, {Genzel}, {Ivison}, {Le Floc'h}, {Magdis},
  {Maiolino}, {Nordon}, {Oliver}, {Page}, {P{\'e}rez Garc{\'{\i}}a},
  {Poglitsch}, {Popesso}, {Pozzi}, {Riguccini}, {Rodighiero}, {Rosario},
  {Roseboom}, {Sanchez-Portal}, {Scott}, {Sturm}, {Tacconi}, {Valtchanov},
  {Wang}, \& {Wuyts}}]{magne14}
---. 2014, \aap, 561, A86

\bibitem[{{Mao} {et~al.}(2011){Mao}, {Huynh}, {Norris}, {Dickinson}, {Frayer},
  {Helou}, \& {Monkiewicz}}]{mao11}
{Mao}, M.~Y., {Huynh}, M.~T., {Norris}, R.~P., {Dickinson}, M., {Frayer}, D.,
  {Helou}, G., \& {Monkiewicz}, J.~A. 2011, \apj, 731, 79

\bibitem[{{Miller} {et~al.}(2013){Miller}, {Bonzini}, {Fomalont}, {Kellermann},
  {Mainieri}, {Padovani}, {Rosati}, {Tozzi}, \& {Vattakunnel}}]{mille13}
{Miller}, N.~A., {et~al.} 2013, \apjs, 205, 13

\bibitem[{{Morrison} {et~al.}(2010){Morrison}, {Owen}, {Dickinson}, {Ivison},
  \& {Ibar}}]{morri10}
{Morrison}, G.~E., {Owen}, F.~N., {Dickinson}, M., {Ivison}, R.~J., \& {Ibar},
  E. 2010, \apjs, 188, 178

\bibitem[{{Murphy}(2009)}]{murph09}
{Murphy}, E.~J. 2009, \apj, 706, 482

\bibitem[{{Newman} {et~al.}(2013){Newman}, {Cooper}, {Davis}, {Faber}, {Coil},
  {Guhathakurta}, {Koo}, {Phillips}, {Conroy}, {Dutton}, {Finkbeiner}, {Gerke},
  {Rosario}, {Weiner}, {Willmer}, {Yan}, {Harker}, {Kassin}, {Konidaris},
  {Lai}, {Madgwick}, {Noeske}, {Wirth}, {Connolly}, {Kaiser}, {Kirby},
  {Lemaux}, {Lin}, {Lotz}, {Luppino}, {Marinoni}, {Matthews}, {Metevier}, \&
  {Schiavon}}]{newma13}
{Newman}, J.~A., {et~al.} 2013, \apjs, 208, 5

\bibitem[{{Niklas} \& {Beck}(1997)}]{nikla97b}
{Niklas}, S., \& {Beck}, R. 1997, \aap, 320, 54

\bibitem[{{Niklas} {et~al.}(1997){Niklas}, {Klein}, \&
  {Wielebinski}}]{nikla97a}
{Niklas}, S., {Klein}, U., \& {Wielebinski}, R. 1997, \aap, 322, 19

\bibitem[{{Norris} {et~al.}(2011){Norris}, {Hopkins}, {Afonso}, {Brown},
  {Condon}, {Dunne}, {Feain}, {Hollow}, {Jarvis}, {Johnston-Hollitt}, {Lenc},
  {Middelberg}, {Padovani}, {Prandoni}, {Rudnick}, {Seymour}, {Umana},
  {Andernach}, {Alexander}, {Appleton}, {Bacon}, {Banfield}, {Becker}, {Brown},
  {Ciliegi}, {Jackson}, {Eales}, {Edge}, {Gaensler}, {Giovannini}, {Hales},
  {Hancock}, {Huynh}, {Ibar}, {Ivison}, {Kennicutt}, {Kimball}, {Koekemoer},
  {Koribalski}, {L{\'o}pez-S{\'a}nchez}, {Mao}, {Murphy}, {Messias},
  {Pimbblet}, {Raccanelli}, {Randall}, {Reiprich}, {Roseboom},
  {R{\"o}ttgering}, {Saikia}, {Sharp}, {Slee}, {Smail}, {Thompson}, {Urquhart},
  {Wall}, \& {Zhao}}]{norri11}
{Norris}, R.~P., {et~al.} 2011, PASA, 28, 215

\bibitem[{{Oliver} {et~al.}(2012){Oliver}, {Bock}, {Altieri}, {Amblard},
  {Arumugam}, {Aussel}, {Babbedge}, {Beelen}, {B{\'e}thermin}, {Blain},
  {Boselli}, {Bridge}, {Brisbin}, {Buat}, {Burgarella},
  {Castro-Rodr{\'{\i}}guez}, {Cava}, {Chanial}, {Cirasuolo}, {Clements},
  {Conley}, {Conversi}, {Cooray}, {Dowell}, {Dubois}, {Dwek}, {Dye}, {Eales},
  {Elbaz}, {Farrah}, {Feltre}, {Ferrero}, {Fiolet}, {Fox}, {Franceschini},
  {Gear}, {Giovannoli}, {Glenn}, {Gong}, {Gonz{\'a}lez Solares}, {Griffin},
  {Halpern}, {Harwit}, {Hatziminaoglou}, {Heinis}, {Hurley}, {Hwang}, {Hyde},
  {Ibar}, {Ilbert}, {Isaak}, {Ivison}, {Lagache}, {Le Floc'h}, {Levenson},
  {Faro}, {Lu}, {Madden}, {Maffei}, {Magdis}, {Mainetti}, {Marchetti},
  {Marsden}, {Marshall}, {Mortier}, {Nguyen}, {O'Halloran}, {Omont}, {Page},
  {Panuzzo}, {Papageorgiou}, {Patel}, {Pearson}, {P{\'e}rez-Fournon}, {Pohlen},
  {Rawlings}, {Raymond}, {Rigopoulou}, {Riguccini}, {Rizzo}, {Rodighiero},
  {Roseboom}, {Rowan-Robinson}, {S{\'a}nchez Portal}, {Schulz}, {Scott},
  {Seymour}, {Shupe}, {Smith}, {Stevens}, {Symeonidis}, {Trichas}, {Tugwell},
  {Vaccari}, {Valtchanov}, {Vieira}, {Viero}, {Vigroux}, {Wang}, {Ward},
  {Wardlow}, {Wright}, {Xu}, \& {Zemcov}}]{olive12}
{Oliver}, S.~J., {et~al.} 2012, \mnras, 424, 1614

\bibitem[{{Price} \& {Duric}(1992)}]{price92}
{Price}, R., \& {Duric}, N. 1992, \apj, 401, 81

\bibitem[{{Roseboom} {et~al.}(2010){Roseboom}, {Oliver}, {Kunz}, {Altieri},
  {Amblard}, {Arumugam}, {Auld}, {Aussel}, {Babbedge}, {B{\'e}thermin},
  {Blain}, {Bock}, {Boselli}, {Brisbin}, {Buat}, {Burgarella},
  {Castro-Rodr{\'{\i}}guez}, {Cava}, {Chanial}, {Chapin}, {Clements}, {Conley},
  {Conversi}, {Cooray}, {Dowell}, {Dwek}, {Dye}, {Eales}, {Elbaz}, {Farrah},
  {Fox}, {Franceschini}, {Gear}, {Glenn}, {Solares}, {Griffin}, {Halpern},
  {Harwit}, {Hatziminaoglou}, {Huang}, {Ibar}, {Isaak}, {Ivison}, {Lagache},
  {Levenson}, {Lu}, {Madden}, {Maffei}, {Mainetti}, {Marchetti}, {Marsden},
  {Mortier}, {Nguyen}, {O'Halloran}, {Omont}, {Page}, {Panuzzo},
  {Papageorgiou}, {Patel}, {Pearson}, {P{\'e}rez-Fournon}, {Pohlen},
  {Rawlings}, {Raymond}, {Rigopoulou}, {Rizzo}, {Rowan-Robinson}, {Portal},
  {Schulz}, {Scott}, {Seymour}, {Shupe}, {Smith}, {Stevens}, {Symeonidis},
  {Trichas}, {Tugwell}, {Vaccari}, {Valtchanov}, {Vieira}, {Vigroux}, {Wang},
  {Ward}, {Wright}, {Xu}, \& {Zemcov}}]{roseb10}
{Roseboom}, I.~G., {et~al.} 2010, \mnras, 409, 48

\bibitem[{{Roychowdhury} \& {Chengalur}(2012)}]{roych12}
{Roychowdhury}, S., \& {Chengalur}, J.~N. 2012, \mnras, 423, L127

\bibitem[{{Sanders} \& {Mirabel}(1996)}]{sande96}
{Sanders}, D.~B., \& {Mirabel}, I.~F. 1996, \araa, 34, 749

\bibitem[{{Sargent} {et~al.}(2010){Sargent}, {Schinnerer}, {Murphy}, {Aussel},
  {Le Floc'h}, {Frayer}, {Mart{\'{\i}}nez-Sansigre}, {Oesch}, {Salvato},
  {Smol{\v c}i{\'c}}, {Zamorani}, {Brusa}, {Cappelluti}, {Carilli}, {Carollo},
  {Ilbert}, {Kartaltepe}, {Koekemoer}, {Lilly}, {Sanders}, \&
  {Scoville}}]{sarge10}
{Sargent}, M.~T., {et~al.} 2010, \apjs, 186, 341

\bibitem[{{Schinnerer} {et~al.}(2010){Schinnerer}, {Sargent}, {Bondi}, {Smol{\v
  c}i{\'c}}, {Datta}, {Carilli}, {Bertoldi}, {Blain}, {Ciliegi}, {Koekemoer},
  \& {Scoville}}]{schinn10}
{Schinnerer}, E., {et~al.} 2010, \apjs, 188, 384

\bibitem[{{Schleicher} \& {Beck}(2013)}]{schle13}
{Schleicher}, D.~R.~G., \& {Beck}, R. 2013, \aap, 556, A142

\bibitem[{{Simpson} {et~al.}(2006){Simpson}, {Mart{\'{\i}}nez-Sansigre},
  {Rawlings}, {Ivison}, {Akiyama}, {Sekiguchi}, {Takata}, {Ueda}, \&
  {Watson}}]{simps06}
{Simpson}, C., {et~al.} 2006, \mnras, 372, 741

\bibitem[{{Skibba} {et~al.}(2014){Skibba}, {Smith}, {Coil}, {Moustakas},
  {Aird}, {Blanton}, {Bray}, {Cool}, {Eisenstein}, {Mendez}, {Wong}, \&
  {Zhu}}]{skibb14}
{Skibba}, R.~A., {et~al.} 2014, \apj, 784, 128

\bibitem[{{Smith} {et~al.}(2012){Smith}, {Wang}, {Oliver}, {Auld}, {Bock},
  {Brisbin}, {Burgarella}, {Chanial}, {Chapin}, {Clements}, {Conversi},
  {Cooray}, {Dowell}, {Eales}, {Farrah}, {Franceschini}, {Glenn}, {Griffin},
  {Ivison}, {Mortier}, {Page}, {Papageorgiou}, {Pearson}, {P{\'e}rez-Fournon},
  {Pohlen}, {Rawlings}, {Raymond}, {Rodighiero}, {Roseboom}, {Rowan-Robinson},
  {Savage}, {Scott}, {Seymour}, {Symeonidis}, {Tugwell}, {Vaccari},
  {Valtchanov}, {Vigroux}, {Ward}, {Wright}, \& {Zemcov}}]{smith12}
{Smith}, A.~J., {et~al.} 2012, \mnras, 419, 377

\bibitem[{{Smith} {et~al.}(2014){Smith}, {Jarvis}, {Hardcastle}, {Vaccari},
  {Bourne}, {Dunne}, {Ibar}, {Maddox}, {Prescott}, {Vlahakis}, {Eales},
  {Maddox}, {Smith}, {Valiante}, \& {de Zotti}}]{smith14}
{Smith}, D.~J.~B., {et~al.} 2014, \mnras, 445, 2232

\bibitem[{{Taylor} {et~al.}(2015){Taylor}, {Hopkins}, {Baldry},
  {Bland-Hawthorn}, {Brown}, {Colless}, {Driver}, {Norberg}, {Robotham},
  {Alpaslan}, {Brough}, {Cluver}, {Gunawardhana}, {Kelvin}, {Liske},
  {Conselice}, {Croom}, {Foster}, {Jarrett}, {Lara-Lopez}, \&
  {Loveday}}]{taylo15}
{Taylor}, E.~N., {et~al.} 2015, \mnras, 446, 2144

\bibitem[{{Tojeiro} {et~al.}(2013){Tojeiro}, {Masters}, {Richards}, {Percival},
  {Bamford}, {Maraston}, {Nichol}, {Skibba}, \& {Thomas}}]{tojei13}
{Tojeiro}, R., {et~al.} 2013, \mnras, 432, 359

\bibitem[{{Veilleux} {et~al.}(2002){Veilleux}, {Kim}, \& {Sanders}}]{veill02}
{Veilleux}, S., {Kim}, D.-C., \& {Sanders}, D.~B. 2002, \apjs, 143, 315

\bibitem[{{V{\"o}lk}(1989)}]{volk89}
{V{\"o}lk}, H.~J. 1989, \aap, 218, 67

\bibitem[{{White} {et~al.}(2007){White}, {Helfand}, {Becker}, {Glikman}, \& {de
  Vries}}]{white07}
{White}, R.~L., {Helfand}, D.~J., {Becker}, R.~H., {Glikman}, E., \& {de
  Vries}, W. 2007, \apj, 654, 99

\bibitem[{{Wunderlich} {et~al.}(1987){Wunderlich}, {Wielebinski}, \&
  {Klein}}]{wunde87}
{Wunderlich}, E., {Wielebinski}, R., \& {Klein}, U. 1987, \aaps, 69, 487

\bibitem[{{Wyder} {et~al.}(2007){Wyder}, {Martin}, {Schiminovich}, {Seibert},
  {Budav{\'a}ri}, {Treyer}, {Barlow}, {Forster}, {Friedman}, {Morrissey},
  {Neff}, {Small}, {Bianchi}, {Donas}, {Heckman}, {Lee}, {Madore}, {Milliard},
  {Rich}, {Szalay}, {Welsh}, \& {Yi}}]{wyder07}
{Wyder}, T.~K., {et~al.} 2007, \apjs, 173, 293

\bibitem[{{Yun} {et~al.}(2001){Yun}, {Reddy}, \& {Condon}}]{yun01}
{Yun}, M.~S., {Reddy}, N.~A., \& {Condon}, J.~J. 2001, \apj, 554, 803

\bibitem[{{Zhu} {et~al.}(2011){Zhu}, {Blanton}, {Burles}, {Coil}, {Cool},
  {Eisenstein}, {Moustakas}, {Wong}, \& {Aird}}]{zhu11}
{Zhu}, G., {et~al.} 2011, \apj, 726, 110



\end{thebibliography}

\appendix

\section{Stacking analysis}

Results of our stacking analysis for various wavebands are given in
Table~\ref{stacks}.  Column 2 shows the range of redshift bin and column 3
shows the number of PRIMUS blue galaxies in each bin. In column 4 we list the
mean redshift of the galaxies. Columns 5 and 6 lists the mean stacked flux
density and the rms noise of the stacked images in mJy, respectively. In column
7 we list the luminosity derived from stacking in the luminosity space in
units of $10^{22}$ W Hz$^{-1}$.  For comparison, we list the luminosity
derived from stacked flux density and the mean redshift in column 8. We have
used the values from column 7 in our analysis. The errors on mean stacked flux
density and luminosity are the 1$\sigma$ bootstrap errors.

\begin{table*}
 \centering
  \caption{Results of stacking analysis.}
\scriptsize
   \begin{tabular}{@{}lccccccc@{}}
  \hline
  \multicolumn{1}{l}{Survey}&
  \multicolumn{1}{c}{Redshift}&
  \multicolumn{1}{c}{$N_{\rm obj}$}&
 \multicolumn{1}{c}{$z_{\rm mean}$}&
  \multicolumn{1}{c}{Mean flux}&
  \multicolumn{1}{c}{Stack image}&
  \multicolumn{2}{c}{Luminosity ($\rm \times 10^{22}~W~Hz^{-1}$)}\\
  \cline{7-8}
	     &  bin    &  &   &density (mJy)  & rms (mJy) & Stacked & Mean \\
	(1)     &  (2)    & (3) & (4)  & (5)  & (6) & (7) & (8) \\
 \hline

 0.325 GHz GMRT        &  0--0.1 &  1525  & 0.064  & 0.027$\pm$0.020  & 0.0088 & 0.06$\pm$0.04 &  0.03$\pm$0.02 \\
				     &  0.1--0.2 &  2378  & 0.155  & 0.024$\pm$0.006  & 0.0052 & 0.18$\pm$0.04 &  0.16$\pm$0.04 \\
		&  0.2--0.3 &  2836  & 0.252  & 0.037$\pm$0.007  & 0.0055 & 0.71$\pm$0.15 &  0.74$\pm$0.13 \\
  &  0.3--0.4 &  3268  & 0.349  & 0.029$\pm$0.002  & 0.0058 & 1.24$\pm$0.18 &  1.23$\pm$0.10 \\
  &  0.4--0.5 &  4410  & 0.451  & 0.031$\pm$0.004  & 0.0042 & 2.37$\pm$0.29 &  2.36$\pm$0.31 \\
  &  0.5--0.6 &  4476  & 0.552  & 0.030$\pm$0.005  & 0.0038 & 3.88$\pm$0.73  &  3.74$\pm$0.58\\
  &  0.6--0.7 &  5339  & 0.646  & 0.030$\pm$0.005  & 0.0041 & 5.53$\pm$0.56 &  5.50$\pm$0.97 \\
  &  0.7--0.8 &  4374  & 0.749  & 0.034$\pm$0.005  & 0.0043 & 8.79$\pm$0.85 &  8.86$\pm$1.23 \\
  &  0.8--0.9 &  3548  & 0.848  & 0.035$\pm$0.009  & 0.0035 & 12.41$\pm$3.38 &  12.54$\pm$3.26 \\
  &  0.9--1.0 &  2114  & 0.944  & 0.046$\pm$0.013  & 0.0059 & 20.98$\pm$4.96 &  21.46$\pm$6.29 \\
  &  1.0--1.1 &  1348  & 1.047  & 0.046$\pm$0.013  & 0.0067 & 27.21$\pm$9.04 &  27.68$\pm$8.11 \\
  &  1.1--1.2 &  1160  & 1.145  & 0.046$\pm$0.011  & 0.0072 & 36.10$\pm$10.02 &  35.09$\pm$8.71 \\
\hline
1.4 GHz VLA         &  0--0.1   &  722  & 0.064  & 0.017$\pm$0.005  & 0.0026 & 0.01$\pm$0.00 &  0.02$\pm$0.00 \\
				  &  0.1--0.2 &  1251  & 0.156  & 0.012$\pm$0.005  & 0.0018 & 0.08$\pm$0.04 &  0.08$\pm$0.03 \\
      &  0.2--0.3 &  1384  & 0.250  & 0.013$\pm$0.002  & 0.0019 & 0.27$\pm$0.07 &  0.25$\pm$0.04 \\
      &  0.3--0.4 &  1639  & 0.350  & 0.010$\pm$0.001  & 0.0015 & 0.43$\pm$0.05 &  0.43$\pm$0.06 \\
      &  0.4--0.5 &  2450  & 0.451  & 0.008$\pm$0.003  & 0.0014 & 0.65$\pm$0.09 &  0.64$\pm$0.22 \\
      &  0.5--0.6 &  2664  & 0.552  & 0.009$\pm$0.002  & 0.0013 & 1.12$\pm$0.33 &  1.13$\pm$0.19 \\
      &  0.6--0.7 &  3040  & 0.647  & 0.012$\pm$0.003  & 0.0012 & 2.25$\pm$0.59 &  2.25$\pm$0.47 \\
      &  0.7--0.8 &  2270  & 0.749  & 0.009$\pm$0.009  & 0.0029 & 2.26$\pm$1.97 &  2.27$\pm$2.37 \\
      &  0.8--0.9 &  1941  & 0.848  & 0.011$\pm$0.002  & 0.0014 & 4.14$\pm$0.76 &  4.12$\pm$0.65 \\
      &  0.9--1.0 &  1079  & 0.943  & 0.011$\pm$0.002  & 0.0028 & 5.16$\pm$1.54 &  5.11$\pm$1.17 \\
      &  1.0--1.1 &  692  & 1.048  & 0.012$\pm$0.003  & 0.0029 & 7.67$\pm$1.86  &  7.54$\pm$1.66 \\
      &  1.1--1.2 &  579  & 1.145  & 0.010$\pm$0.006  & 0.0029 & 7.93$\pm$3.93  &  7.92$\pm$4.32 \\
\hline
24$\mu$m SWIRE         &  0--0.1   &  1524  & 0.064  &  0.027$\pm$0.004  &  0.0016 & 0.04$\pm$0.005 &   0.03$\pm$0.004 \\
				     &  0.1--0.2 &  2370  & 0.155  &  0.062$\pm$0.004  &  0.0012 & 0.46$\pm$0.03 &  0.41$\pm$0.02 \\
		&  0.2--0.3 &  2829  & 0.252  &  0.077$\pm$0.004  &  0.0013 & 1.57$\pm$0.10  &  1.51$\pm$0.08 \\
  &  0.3--0.4 &  3263  & 0.349  &  0.070$\pm$0.002  &  0.0013 & 2.88$\pm$0.16 &  2.93$\pm$0.08 \\
  &  0.4--0.5 &  4403  & 0.451  &  0.054$\pm$0.001  &  0.0012 & 4.12$\pm$0.16 &  4.16$\pm$0.11 \\
  &  0.5--0.6 &  4471  & 0.552  &  0.050$\pm$0.002  &  0.0010 & 6.23$\pm$0.36 &  6.22$\pm$0.25 \\
  &  0.6--0.7 &  5325  & 0.646  &  0.048$\pm$0.002  &  0.0009 & 8.92$\pm$0.38 &  8.85$\pm$0.34 \\
  &  0.7--0.8 &  4352  & 0.749  &  0.058$\pm$0.002  &  0.0012 & 15.32$\pm$0.51 &  15.31$\pm$0.43 \\
  &  0.8--0.9 &  3539  & 0.848  &  0.056$\pm$0.002  &  0.0010 & 20.27$\pm$0.56 &  20.22$\pm$0.68 \\
  &  0.9--1.0 &  2109  & 0.944  &  0.064$\pm$0.001  &  0.0021 & 30.32$\pm$1.04 &  30.16$\pm$0.69 \\
  &  1.0--1.1 &  1345  & 1.047  &  0.059$\pm$0.003  &  0.0023 & 35.86$\pm$1.73 &  35.91$\pm$2.04 \\
  &  1.1--1.2 &  1158  & 1.145  &  0.044$\pm$0.003 &  0.0022 & 33.57$\pm$1.93 &  33.55$\pm$2.48 \\

\hline
$70\mu$m SWIRE        &  0--0.1 &  1519  & 0.064  &  0.469$\pm$0.080  &  0.0963 & 0.64$\pm$0.12 &  0.46$\pm$0.08 \\
		&  0.1--0.2 &  2365  & 0.155  &  0.913$\pm$0.058  &  0.0871 & 6.57$\pm$0.49  &  6.00$\pm$0.38 \\
	       &  0.2--0.3 &  2824  & 0.252  &  1.028$\pm$0.088  &  0.0709 & 19.87$\pm$1.76 &  20.19$\pm$1.72 \\
	       &  0.3--0.4 &  3257  & 0.349  &  0.821$\pm$0.053  &  0.0553 & 34.10$\pm$2.15 &  34.23$\pm$2.22 \\
	       &  0.4--0.5 &  4394  & 0.451  &  0.658$\pm$0.056  &  0.0580 & 49.87$\pm$2.79 &  50.41$\pm$4.25 \\
 &  0.5--0.6 &  4457  & 0.552  &  0.571$\pm$0.034  &  0.0600 & 71.30$\pm$6.20 &  71.28$\pm$4.27 \\
 &  0.6--0.7 &  5321  & 0.646  &  0.538$\pm$0.067  &  0.0517 & 98.21$\pm$5.50 &  98.50$\pm$12.17 \\
 &  0.7--0.8 &  4340  & 0.749  &  0.576$\pm$0.040  &  0.0672 & 152.96$\pm$13.28 &  152.22$\pm$10.51 \\
 &  0.8--0.9 &  3534  & 0.848  &  0.401$\pm$0.058  &  0.0538 & 143.00$\pm$15.29 &  144.13$\pm$20.83 \\
 &  0.9--1.0 &  2103  & 0.944  &  0.809$\pm$0.308  &  0.0784 & 377.15$\pm$131.50 &  379.72$\pm$144.33 \\
 &  1.0--1.1 &  1344  & 1.047  &  0.496$\pm$0.094 &  0.0765 & 307.86$\pm$38.37 &  301.52$\pm$57.37 \\
 &  1.1--1.2 &  1156  & 1.145  &  0.309$\pm$0.046  &  0.1277 & 232.92$\pm$43.61 &  234.38$\pm$35.20 \\
\hline
$160\mu$m SWIRE        &  0--0.1   &  1501  & 0.064  &  1.998$\pm$0.387  &  0.5270 & 2.88$\pm$0.46 &  1.95$\pm$0.38 \\
		   &  0.1--0.2 &  2318  & 0.155  &  3.099$\pm$0.368  &  0.4853 & 22.81$\pm$2.23 &  20.30$\pm$2.41 \\
		       &  0.2--0.3 &  2776  & 0.252  &  2.694$\pm$0.242  &  0.3390 & 53.82$\pm$7.23 &  52.88$\pm$4.74 \\
		&  0.3--0.4 &  3190  & 0.349  &  2.981$\pm$0.243  &  0.3311 & 125.05$\pm$10.65  &  124.26$\pm$10.13 \\
  &  0.4--0.5 &  4310  & 0.451  &  2.224$\pm$0.259  &  0.2910 & 167.28$\pm$18.95 &  170.39$\pm$19.83 \\
  &  0.5--0.6 &  4392  & 0.552  &  2.083$\pm$0.199  &  0.3003 & 261.33$\pm$24.83 &  260.11$\pm$24.90 \\
  &  0.6--0.7 &  5223  & 0.646  &  2.003$\pm$0.194  &  0.2349 & 374.07$\pm$38.55 &  366.41$\pm$35.47 \\
  &  0.7--0.8 &  4268  & 0.749  &  1.957$\pm$0.198  &  0.3171 & 526.54$\pm$46.87  &  517.24$\pm$52.21 \\
  &  0.8--0.9 &  3489  & 0.848  &  2.103$\pm$0.195  &  0.4453 & 761.58$\pm$91.49  &  754.85$\pm$69.84 \\
  &  0.9--1.0 &  2074  & 0.944  &  2.165$\pm$0.567  &  0.5286 & 1020.48$\pm$302.42 &  1016.19$\pm$266.30 \\
  &  1.0--1.1 &  1316  & 1.048  &  2.194$\pm$0.321  &  0.6136 & 1372.73$\pm$298.74 &  1334.89$\pm$195.16 \\
  &  1.1--1.2 &  1142  & 1.144  &  1.946$\pm$0.315  &  0.5715 & 1474.78$\pm$371.85 & 1477.01$\pm$239.37  \\
\hline
$250\mu$m HerMES        &  0--0.1   &  1525  & 0.064  & 0.634$\pm$0.139  & 0.2008 & 1.18$\pm$0.22 &  0.62$\pm$0.14   \\
			       &  0.1--0.2 &  2378  & 0.155  & 2.205$\pm$0.183  & 0.1671 & 17.29$\pm$1.43  &  14.49$\pm$1.21   \\
		 &  0.2--0.3 &  2836  & 0.252  & 2.422$\pm$0.224  & 0.1579 & 48.00$\pm$3.96  &  47.60$\pm$4.41   \\
   &  0.3--0.4 &  3268  & 0.349  & 2.691$\pm$0.160  & 0.1440 & 111.75$\pm$6.68 &  112.17$\pm$6.68   \\
   &  0.4--0.5 &  4410  & 0.451  & 2.431$\pm$0.147  & 0.1208 & 186.92$\pm$8.69 &  186.22$\pm$11.23   \\
   &  0.5--0.6 &  4476  & 0.552  & 2.348$\pm$0.146  & 0.1269 & 292.77$\pm$16.40 &  293.08$\pm$18.20   \\
   &  0.6--0.7 &  5339  & 0.646  & 2.347$\pm$0.086  & 0.1219 & 433.28$\pm$21.38 &  429.42$\pm$15.68   \\
   &  0.7--0.8 &  4374  & 0.749  & 2.713$\pm$0.131  & 0.1324 & 714.79$\pm$34.14 &  717.13$\pm$34.72   \\
   &  0.8--0.9 &  3548  & 0.848  & 2.360$\pm$0.146  & 0.1560 & 843.22$\pm$38.97 &  847.67$\pm$52.30   \\
   &  0.9--1.0 &  2114  & 0.944  & 2.913$\pm$0.247  & 0.1885 & 1375.33$\pm$97.87 &  1366.98$\pm$115.95   \\
   &  1.0--1.1 &  1348  & 1.047  & 2.742$\pm$0.203  & 0.2366 & 1680.61$\pm$117.42 &  1667.16$\pm$123.22   \\
   &  1.1--1.2 &  1160  & 1.145  & 2.461$\pm$0.213  & 0.2277 & 1864.39$\pm$166.36 &  1869.03$\pm$161.61   \\
\hline 
\end{tabular}
\end{table*}

\addtocounter{table}{-1}
\begin{table*}
 \centering
 \caption{{\it (continued...)} Results of stacking analysis.}
\scriptsize
   \begin{tabular}{@{}lccccccc@{}}
  \hline
    \multicolumn{1}{l}{Survey}&
  \multicolumn{1}{c}{Redshift}&
  \multicolumn{1}{c}{$N_{\rm obj}$}&
 \multicolumn{1}{c}{$z_{\rm mean}$}&
  \multicolumn{1}{c}{Mean flux}&
  \multicolumn{1}{c}{Stack image}&
  \multicolumn{2}{c}{Luminosity ($\rm \times 10^{22}~W~Hz^{-1}$)}\\
  \cline{7-8}
	     &  bin    &  &   &density (mJy)  & rms (mJy) & Stacked & Mean \\
	(1)     &  (2)    & (3) & (4)  & (5)  & (6) & (7) & (8) \\

\hline
$350\mu$m HerMES        &  0--0.1   &  1525  & 0.064  & 0.471$\pm$0.126  & 0.2139 & 0.77$\pm$0.21  &  0.46$\pm$0.12 \\
					     &  0.1--0.2 &  2378  & 0.155  & 1.038$\pm$0.155  & 0.1972 & 8.02$\pm$0.89 &  6.82$\pm$1.02 \\
	  &  0.2--0.3 &  2836  & 0.252  & 0.818$\pm$0.101  & 0.1886 & 15.64$\pm$1.90  &  16.08$\pm$1.99 \\
   &  0.3--0.4 &  3268  & 0.349  & 1.141$\pm$0.135  & 0.1475 & 45.63$\pm$7.68 &  47.56$\pm$5.62 \\
   &  0.4--0.5 &  4410  & 0.451  & 1.257$\pm$0.098  & 0.1158 & 96.42$\pm$8.35 &  96.30$\pm$7.50 \\
   &  0.5--0.6 &  4476  & 0.552  & 1.233$\pm$0.080  & 0.1266 & 155.87$\pm$15.47 &  153.91$\pm$9.96 \\
   &  0.6--0.7 &  5339  & 0.646  & 1.470$\pm$0.053  & 0.1269 & 272.28$\pm$21.03 &  269.10$\pm$9.71 \\
   &  0.7--0.8 &  4374  & 0.749  & 1.821$\pm$0.145  & 0.1423 & 484.65$\pm$40.00 &  481.47$\pm$38.26 \\
   &  0.8--0.9 &  3548  & 0.848  & 1.781$\pm$0.156  & 0.1591 & 635.81$\pm$46.56  &  639.66$\pm$56.10\\
   &  0.9--1.0 &  2114  & 0.944  & 1.999$\pm$0.224  & 0.1679 & 948.50$\pm$73.44  &  938.25$\pm$105.05 \\
   &  1.0--1.1 &  1348  & 1.047  & 2.386$\pm$0.160  & 0.2496 & 1455.06$\pm$119.08 &  1450.73$\pm$97.26  \\
   &  1.1--1.2 &  1160  & 1.145  & 2.278$\pm$0.268  & 0.2538 & 1730.19$\pm$190.89 &  1729.98$\pm$203.59 \\
\hline
$500\mu$m HerMES   & 0--0.1    &  1525  & 0.064  & 0.599$\pm$0.134  & 0.2326 & 0.87$\pm$0.16 &  0.59$\pm$0.13 \\
				 & 0.1--0.2  &  2378  & 0.155  & 0.368$\pm$0.131  & 0.1971 & 2.88$\pm$0.52 &  2.42$\pm$0.86 \\
     & 0.2--0.3  &  2836  & 0.252  & 0.380$\pm$0.081  & 0.1776 & 9.11$\pm$1.50 &  7.46$\pm$1.60 \\
     & 0.3--0.4  &  3268  & 0.349  & 0.275$\pm$0.099  & 0.1800 & 13.02$\pm$4.35 &  11.46$\pm$4.11 \\
     & 0.4--0.5  &  4410  & 0.451  & 0.451$\pm$0.070  & 0.1235 & 37.97$\pm$9.63 &  34.53$\pm$5.37 \\
     & 0.5--0.6  &  4476  & 0.552  & 0.474$\pm$0.099  & 0.1356 & 59.29$\pm$11.51  &  59.15$\pm$12.41 \\
     & 0.6--0.7  &  5339  & 0.646  & 0.809$\pm$0.080  & 0.1234 & 151.81$\pm$23.11 &  148.13$\pm$14.64 \\
     & 0.7--0.8  &  4374  & 0.749  & 0.941$\pm$0.121  & 0.1507 & 249.93$\pm$40.49  &  248.61$\pm$32.10 \\
     & 0.8--0.9  &  3548  & 0.848  & 0.850$\pm$0.108  & 0.1703 & 298.76$\pm$41.09  &  305.34$\pm$38.78 \\
     & 0.9--1.0  &  2114  & 0.944  & 1.084$\pm$0.192  & 0.1843 & 518.44$\pm$71.76 &  508.75$\pm$89.87 \\
     & 1.0--1.1  &  1348  & 1.047  & 1.288$\pm$0.159  & 0.2314 & 789.56$\pm$122.30 & 783.08$\pm$96.78  \\
     & 1.1--1.2  &  1160  & 1.145  & 1.268$\pm$0.151  & 0.2526 & 960.91$\pm$178.99 &  963.30$\pm$114.79 \\
\hline

\end{tabular}
\label{stacks}
\end{table*}

\section{SED fits to stacked images}

\begin{figure*}
\begin{center}
\begin{tabular}{ccc}
{\mbox{\includegraphics[height=5.0cm]{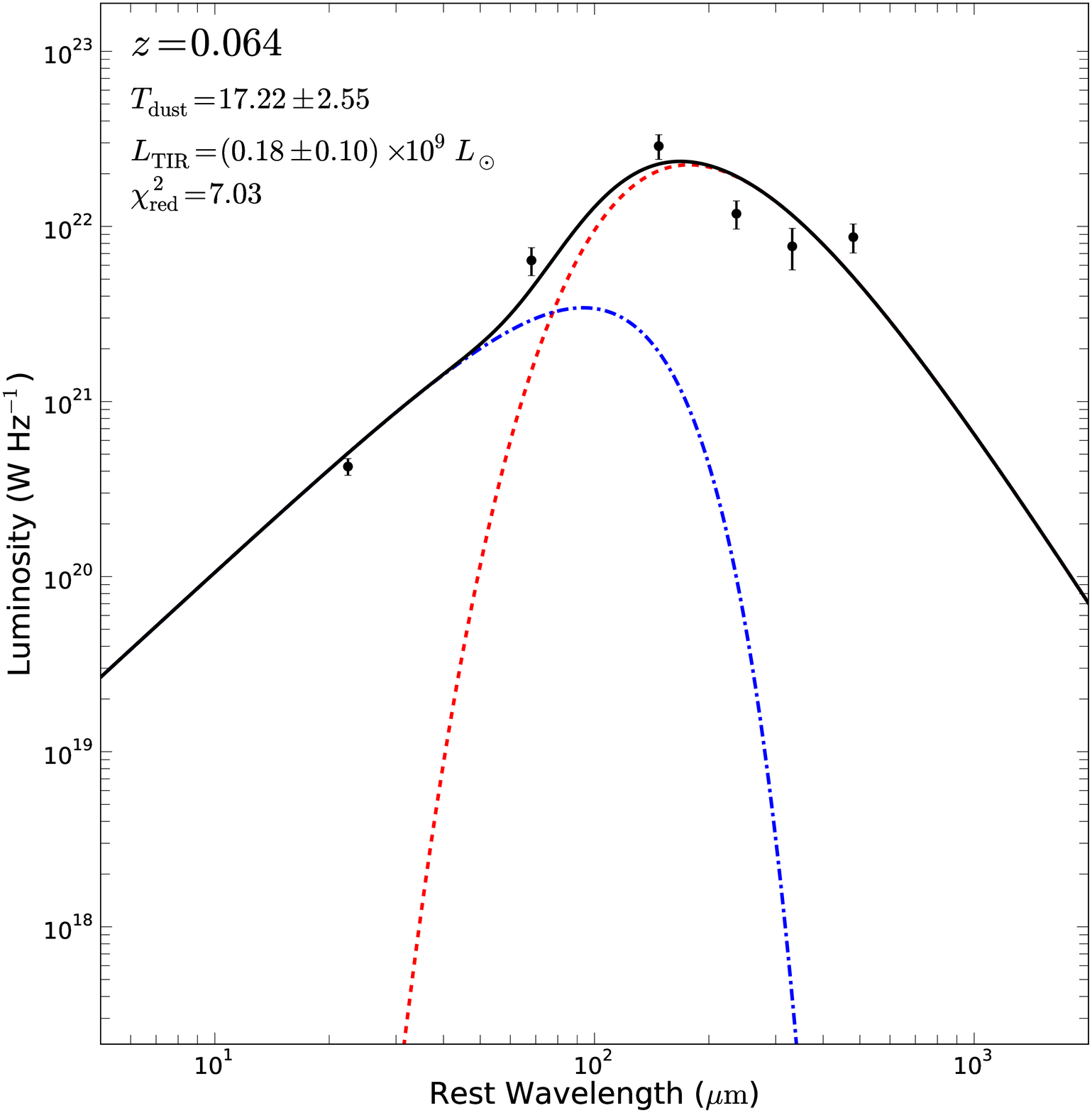}}}&
{\mbox{\includegraphics[height=5.0cm]{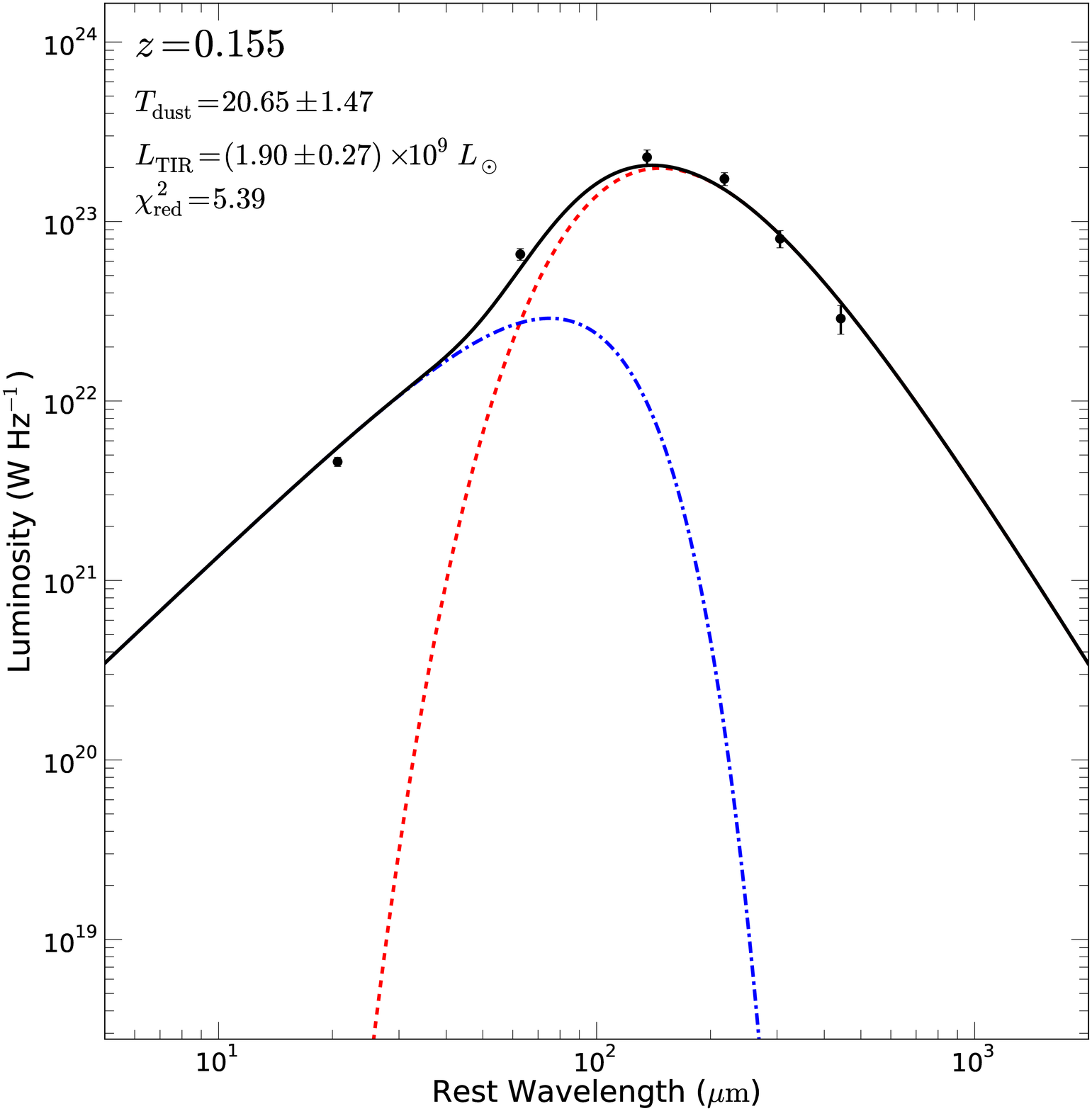}}}&
{\mbox{\includegraphics[height=5.0cm]{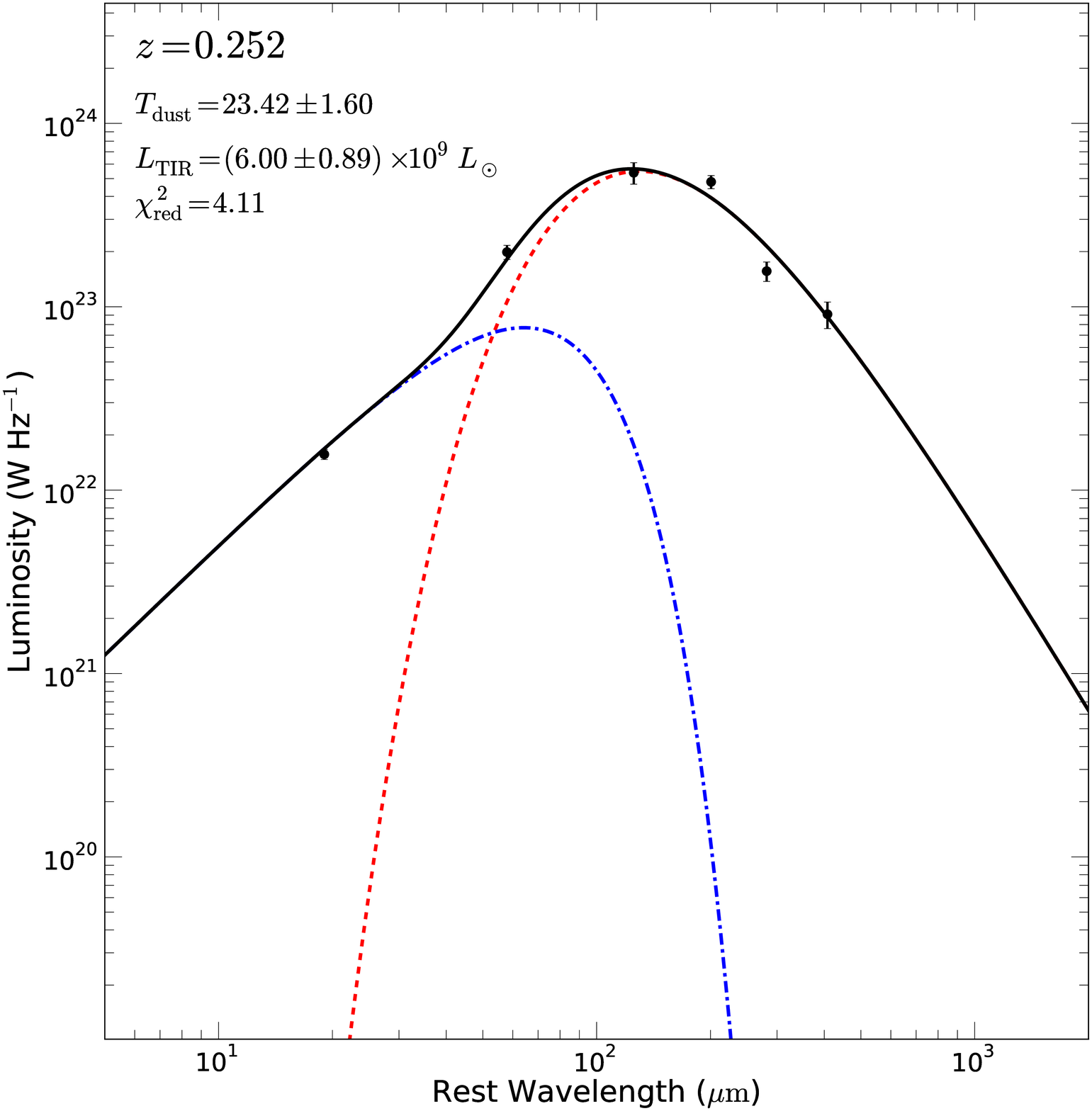}}}\\
{\mbox{\includegraphics[height=5.0cm]{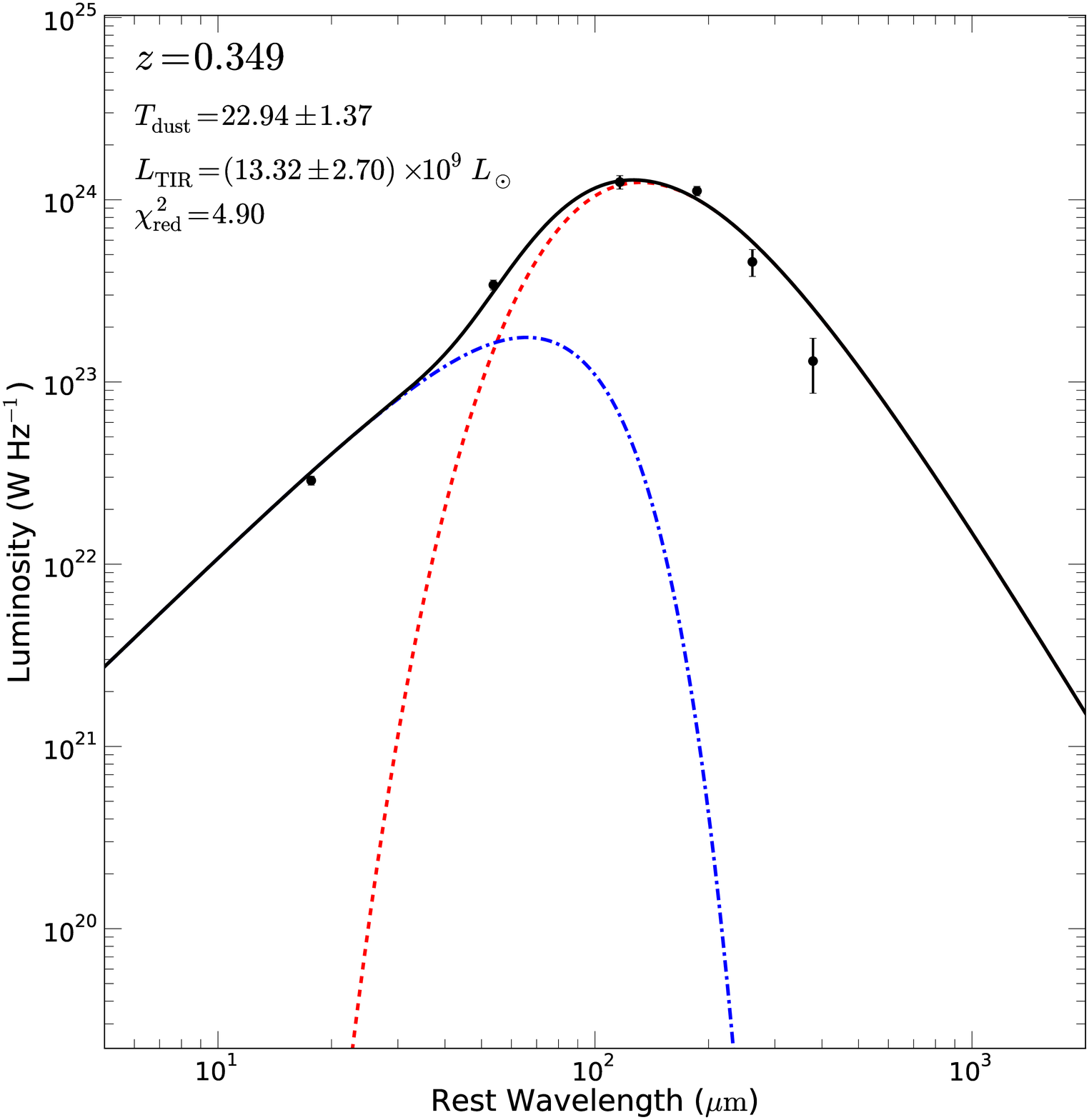}}}&
{\mbox{\includegraphics[height=5.0cm]{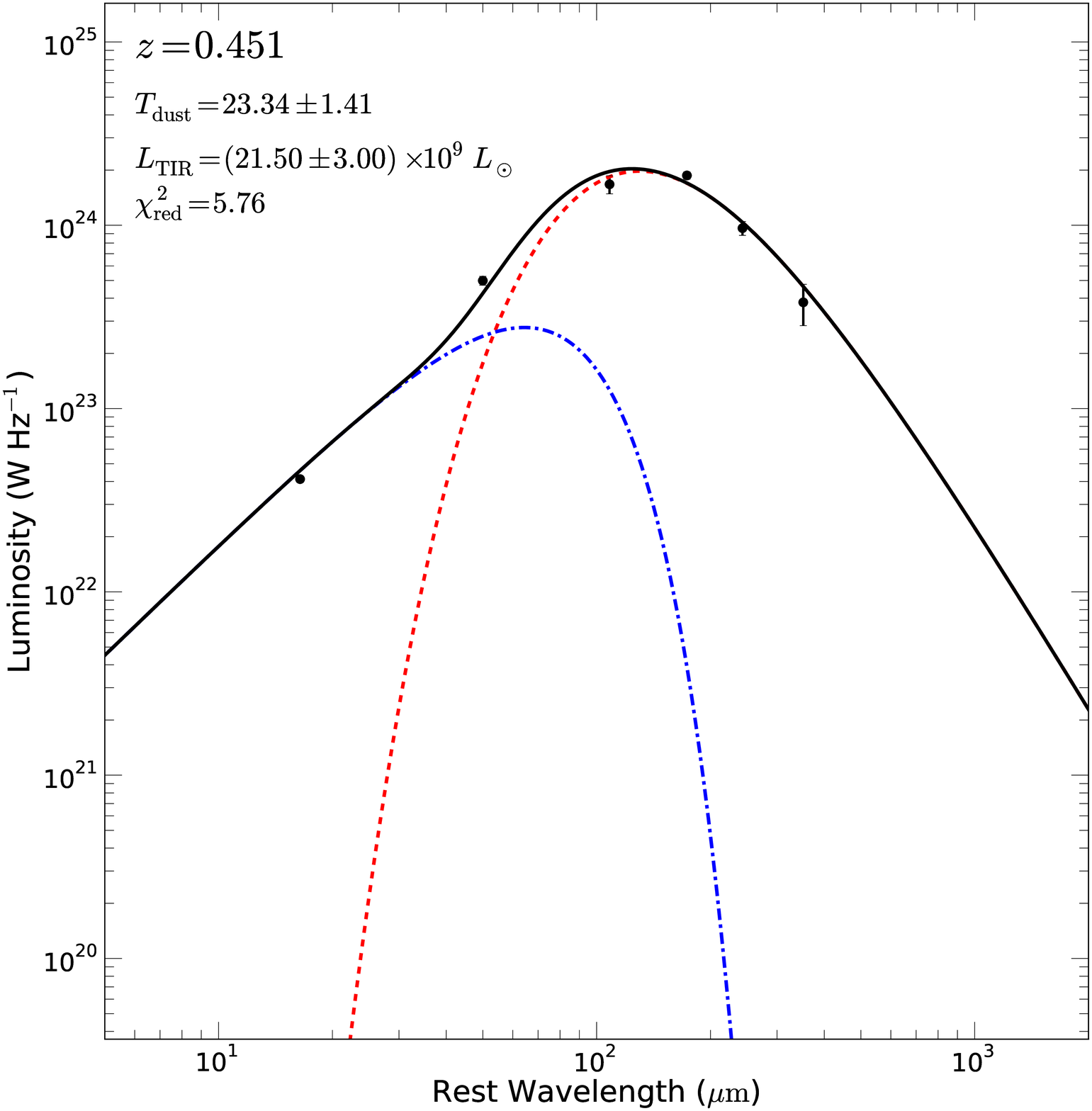}}}&
{\mbox{\includegraphics[height=5.0cm]{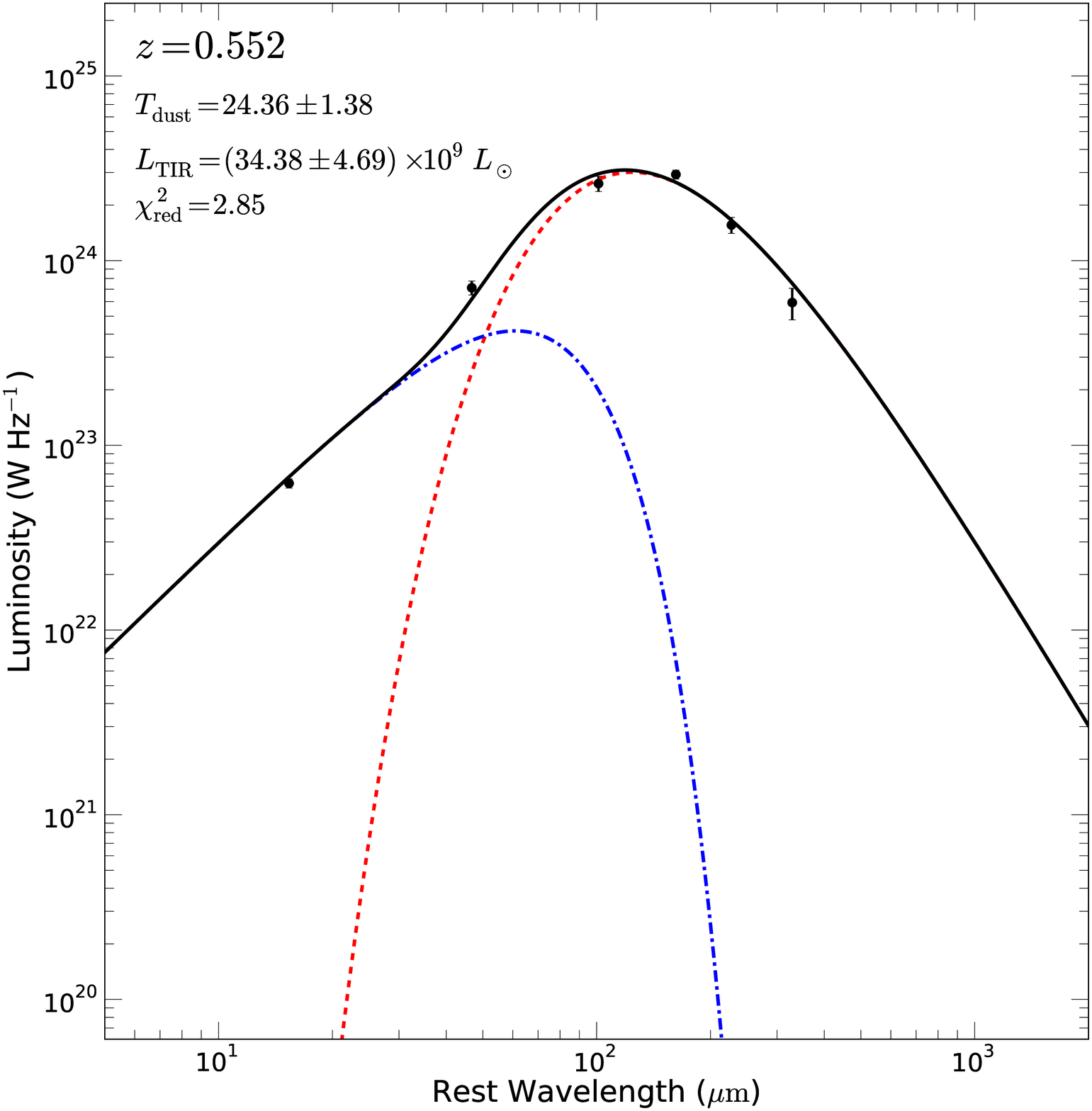}}}\\
{\mbox{\includegraphics[height=5.0cm]{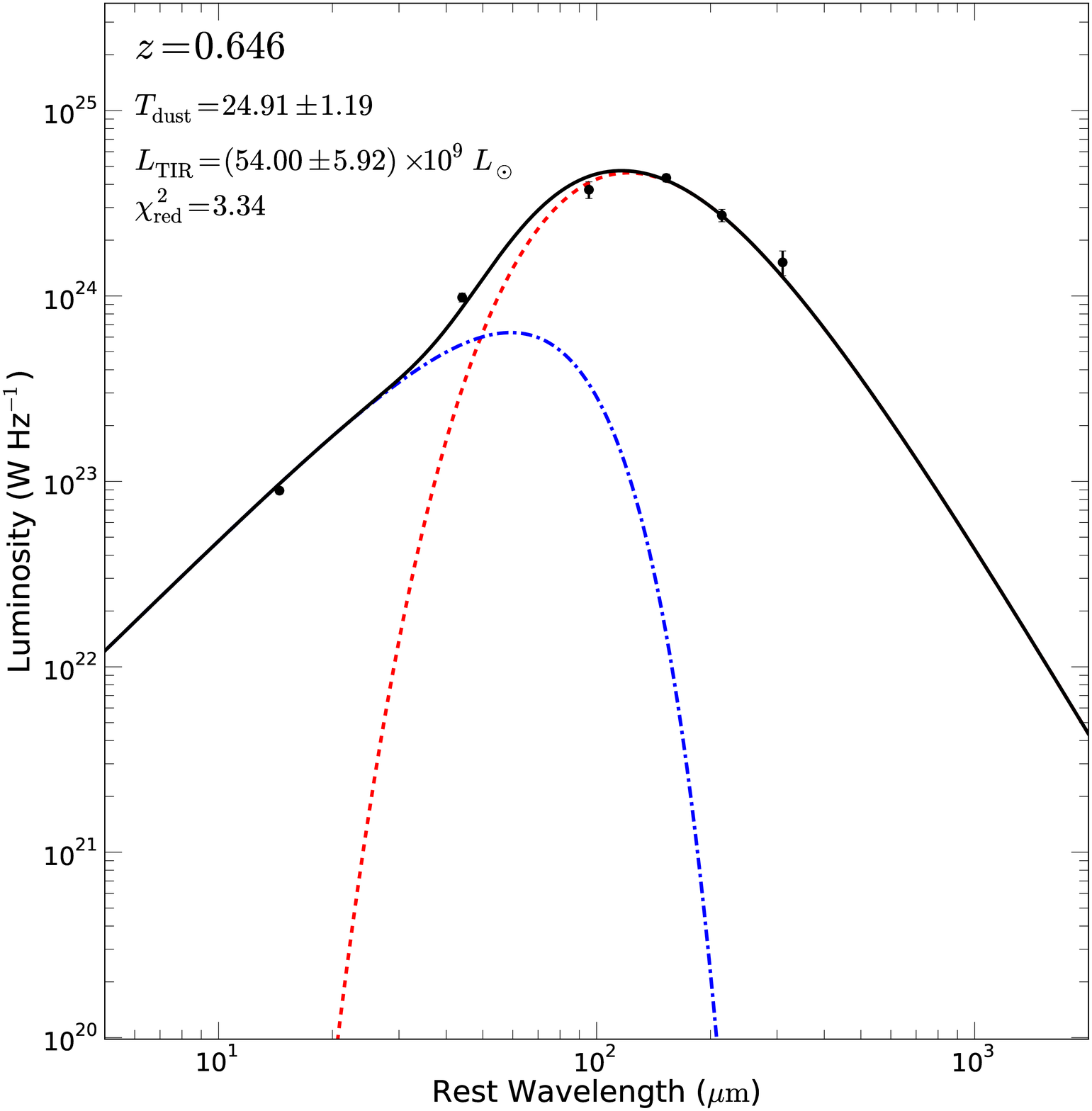}}}&
{\mbox{\includegraphics[height=5.0cm]{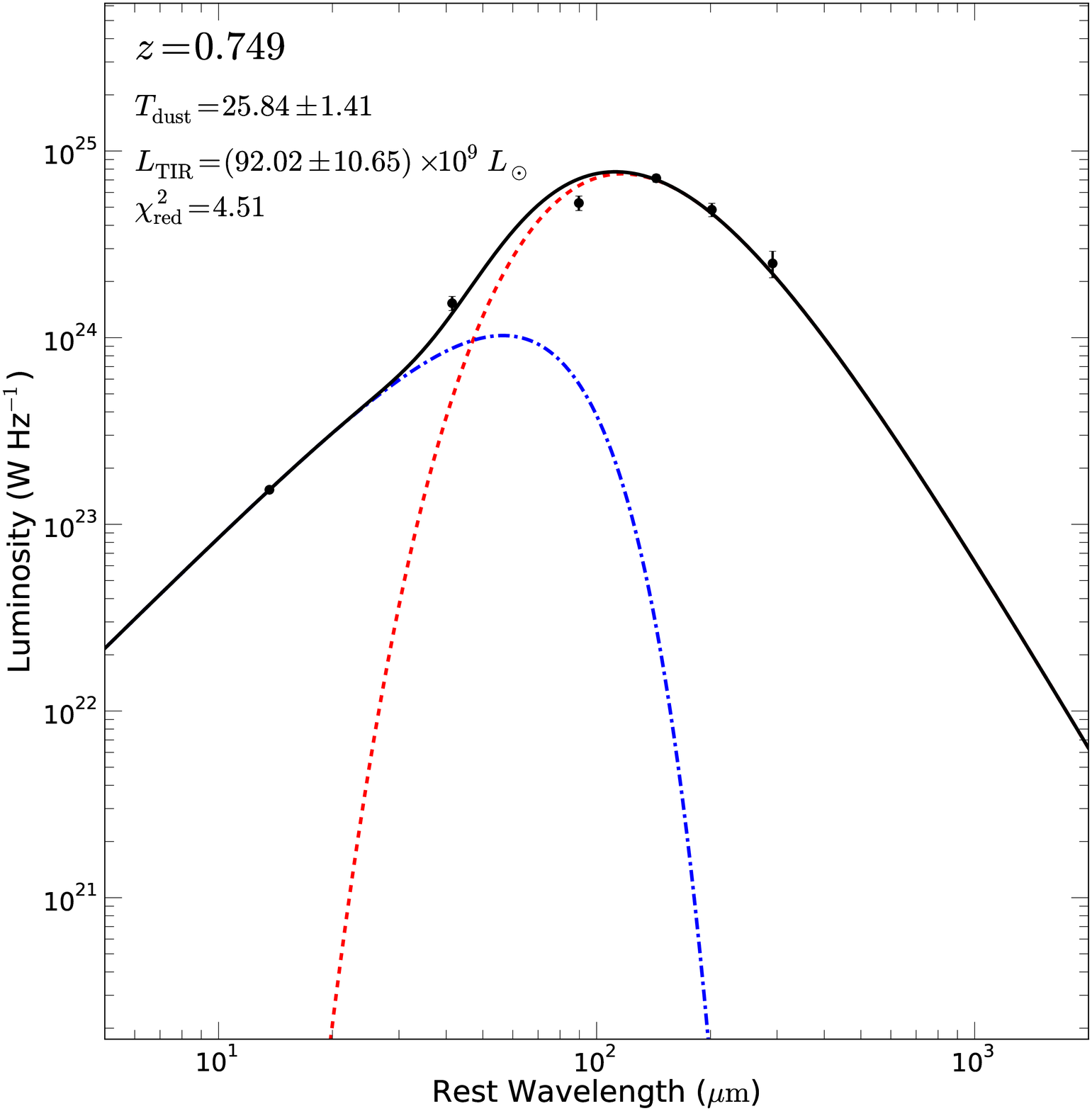}}}&
{\mbox{\includegraphics[height=5.0cm]{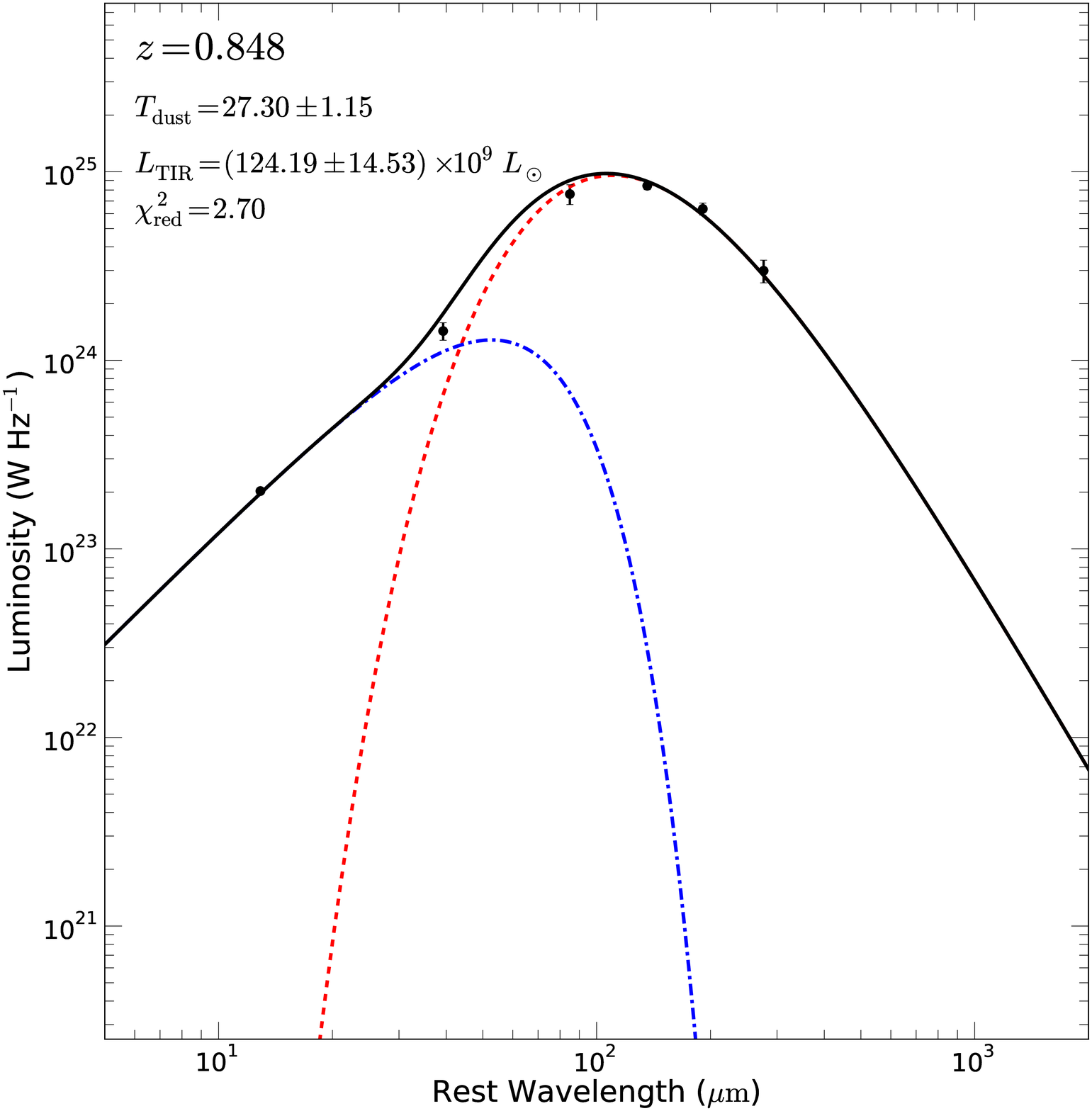}}}\\
{\mbox{\includegraphics[height=5.0cm]{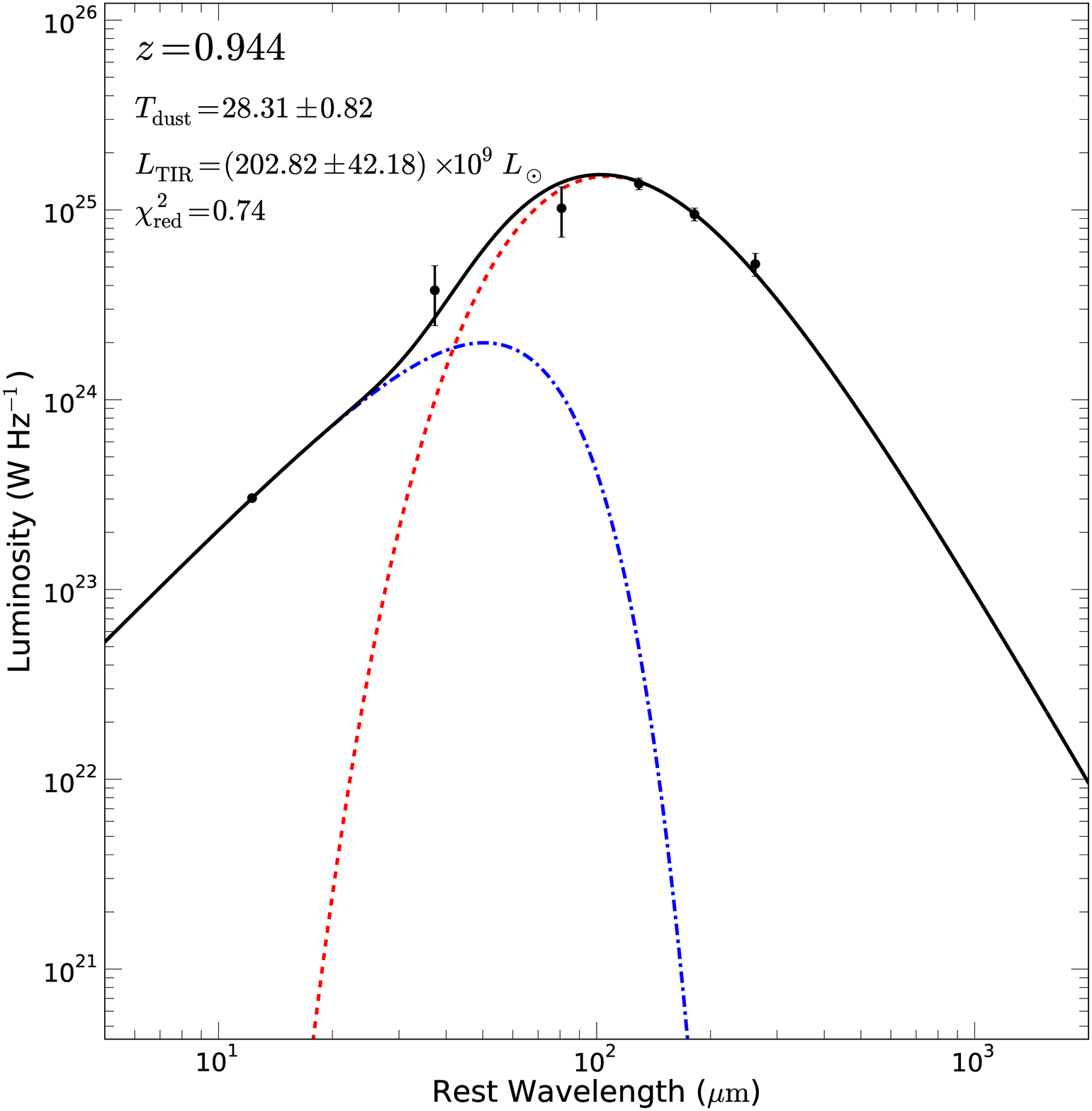}}}&
{\mbox{\includegraphics[height=5.0cm]{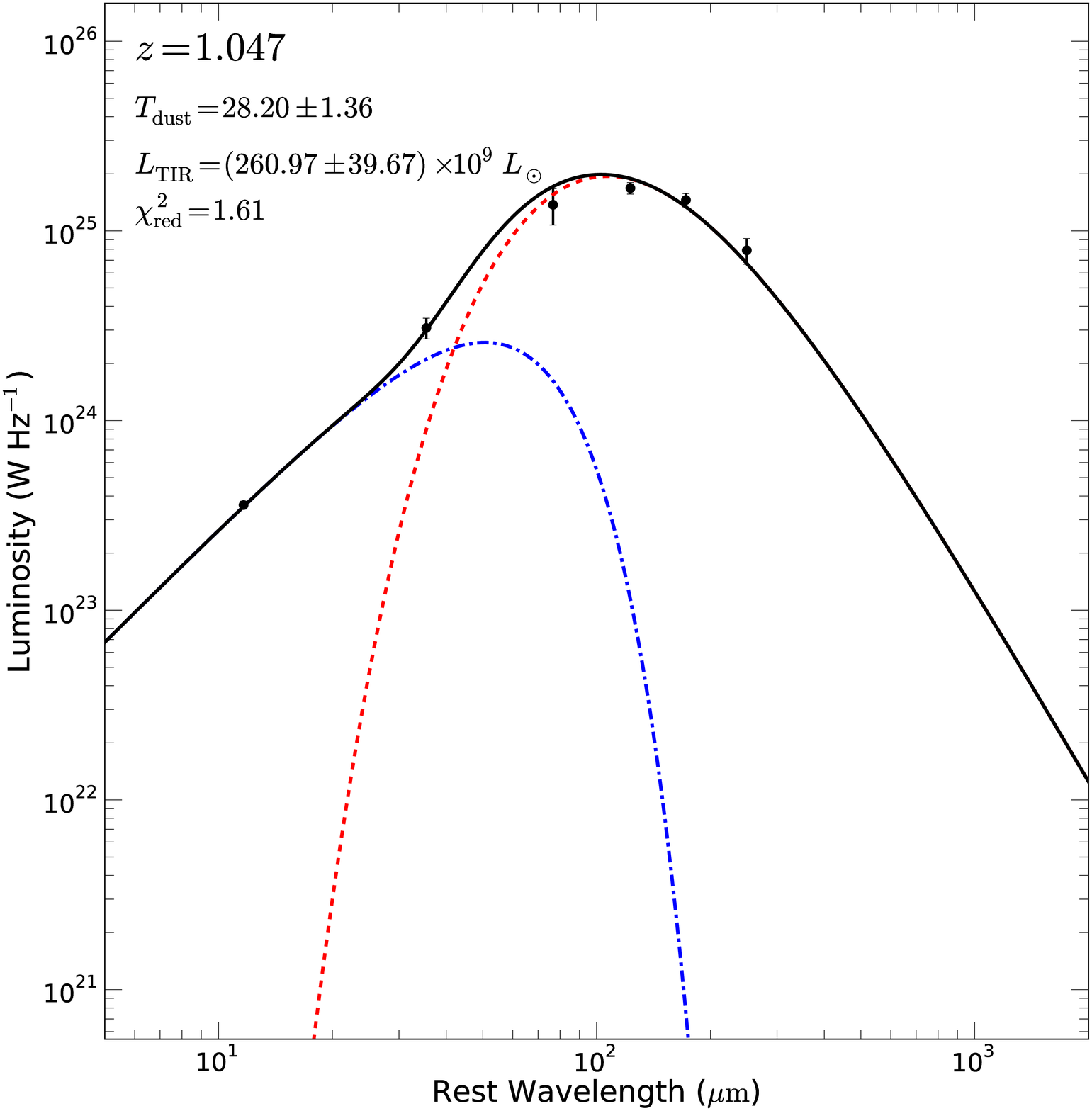}}}&
{\mbox{\includegraphics[height=5.0cm]{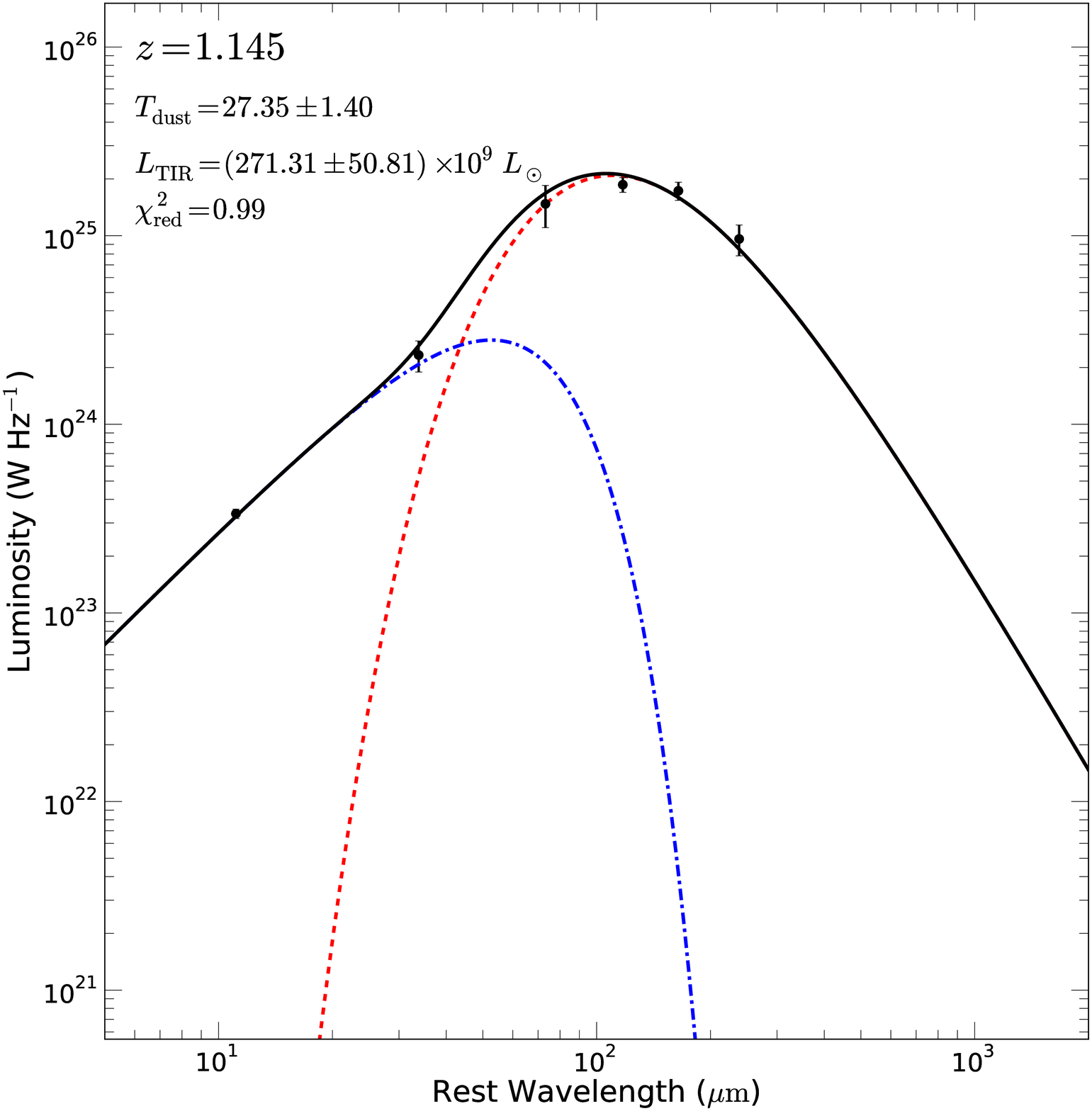}}}\\
\end{tabular}
\end{center}
\caption{Far-infrared SED fit to the stacked images in various redshift
bins.  The red dashed curve represents the modified black body component fit
and the blue dash-dot line represents the mid-infrared power law fit. The black
solid line represents the sum of these components and provides an excellent fit
to the data in almost every case.}
\label{stack-sed-comp}
\end{figure*}

We show the two components of the SED fitting in the FIR regime. In
Figure~\ref{stack-sed-comp}, the greybody component is shown as a dashed red
line and the mid-IR power law component is shown as a blue dashed-dot line. The
solid black curve shows the total fit.

\end{document}